\documentclass[prd,aps,floats,floatfix,eqsecnum,nofootinbib,12pt]{revtex4-1}
\usepackage{verbatim,graphicx,amssymb,amsbsy,bm,amsmath,rotating,epsfig}
\usepackage{psfrag}
\usepackage{hyperref}
\usepackage{fontenc}
\usepackage[latin1]{inputenc}
\usepackage{pstricks,pst-node,pst-text,pst-3d}
\usepackage{textcomp}
\newcommand{\be}{\begin{equation}}
\newcommand{\ee}{\end{equation}}
\newcommand{\bea}{\begin{eqnarray}}
\newcommand{\eea}{\end{eqnarray}}

\begin{document}

\title{Quantum Discrete Levels of the Universe from the Early  Trans-Planckian Vacuum  to the Late Dark Energy}
\author{Norma G. SANCHEZ\\ 
CNRS LERMA Observatoire de Paris PSL University,\\
Sorbonne University, 
61, Avenue de l'Observatoire, 75014 Paris, France}
\date{\today}

\begin{abstract}
{\bf Abstract:} We go forward in completing the standard model of the universe back in time with Planckian and trans-Planckian physics before inflation in agreement with observations, classical-quantum gravity duality and quantum space-time. The quantum vacuum energy bends the space-time and produces a constant curvature de Sitter background. We link de Sitter universe and the cosmological constant to the (classical and quantum) harmonic oscillator.  We find the quantum discrete cosmological levels: size, time, vacuum energy, Hubble constant and gravitational (Gibbons-Hawking) entropy and temperature from the very early trans-Planckian vacuum to the classical today vacuum energy. For each level $n = 0, 1, 2,...$ the two: post and pre (trans)-Planckian phases are covered: In the post-Planckian universe: $ t_{planck} \equiv t_P \leq t \leq 10^{61}t_P$ the levels (in Planck units) are: Hubble constant $H_{n} = {1}/\sqrt{(2n + 1)}$, vacuum energy $\Lambda_{n} = 1/(2n + 1)$, entropy $S_n = (2n + 1)$.  As $n$ increases, radius, mass and $S_n$ increase, $H_n$ and $\Lambda_n$ decrease and {\it consistently} the universe {\it classicalizes}.  In the pre-Planckian (trans-Planckian) phase $10^{-61} t_P \leq t \leq t_P$ the quantum levels are: $H_{Qn} = \sqrt{(2n + 1)},\; \Lambda_{Qn} = (2n + 1),\; S_{Qn} = 1/(2n + 1)$, $Q$ denoting quantum. The $n$-levels cover {\it all} scales from the far past highest excited trans-Planckian level $n = 10^{122}$ with finite curvature, $\Lambda_Q = 10^{122}$ and minimum entropy $S_Q =  10^{-122}$, $n$ decreases till the Planck level $(n = 0)$ with $H_{planck}=1=\Lambda_{planck}= S_{planck}$ and enters the post-Planckian phase e.g. $n = 1, 2,...,n_{inflation}=10^{12},... ,n_{cmb} = 10^{114},...,n_{reoin}=10^{118},...,n_{today} = 10^{122}$ with the most classical value $H_{today} = 10^{-61}$, $\Lambda_{today} = 10^{-122}$, $S_{today}=10^{122}$. We implement the Snyder-Yang algebra in this context yielding a consistent group-theory realization of quantum discrete de Sitter space-time, classical-quantum gravity duality symmetry and a clarifying unifying picture.https://chalonge-devega.fr/sanchez \end{abstract}


\maketitle
\tableofcontents

\section{Introduction and Results}

Planckian and trans-Planckian energies are theoretically allowed, physically 
motivated too, the universe and its very early stages have all the quantum conditions for 
such extreme quantum gravitational regimes and energies, the black hole interiors 
too. The {\it truly} quantum gravity {\it domain} is not reduced to be fixed at the Planck scale or the 
neighborhoods of it, but extends deep beyond the Planck scale in the highly 
quantum trans-Planckian range. 

\medskip

{\bf In this paper} we go forward in completing the standard model of the universe back in time with  Planckian and trans-Planckian physics before inflation in agreement with observations, classical-quantum gravity duality and quantum space-time in this context. 

\medskip
	
Quantum theory is more complete than classical theory and tells us what value a classical observable should have. The classical-quantum (or wave-particle) duality is a robust and universal concept (it does not depend on the nature or number of space-time dimensions, compactified or not, nor on particular space-time geometries, topologies, symmetries, nor on other {\it at priori} condition). Moreover, the quantum trans-Planckian eras in the far past universe determine the post-Planckian eras, e.g. the inflation and the cosmological vacuum energy until today dark energy, namely the evolution from the quantum very early phases to the semi-classical and classical phases and the arrow of time as determined by the gravitational entropy.  

\medskip

The complete universe is composed of two main phases, 
the Planck scale being the {\it transition scale}: the quantum  pre-Planckian or trans-Planckian phase $0 < 10^{-61} t_P \leq t \leq t_P$ and the semiclassical and mostly classical post-Planckian universe $t_P \leq t \leq t_{today}= 10^{61} t_P$, $t_P$ being the Planck time. The pre-Planckian era could be tested indirectly through its post-Planckian observables, e.g. primordial graviton signals, inflation and the CMB till today dark energy. This framework provides in particular the gravitational entropy and temperature (classical, semiclassical and quantum) in the different cosmological regimes and eras \cite{Sanchez2019},\cite{Sanchez3}, in particular the Gibbons-Hawking entropy and temperature. Interesting too (and related with) are the classical and quantum cosmological vacuum energy  values $(\Lambda, \Lambda_Q)$ dual of each other: For instance, the quantum $\Lambda_Q$ obtained from the classical-quantum (or wave-particle) duality approach turns out to be the saddle point obtained from the quantum gravity path integral euclidean approach which action is the well-known Gibbons-Hawking de Sitter entropy, showing the consistency of the results  \cite{Sanchez2019},\cite{Sanchez3}.

\medskip

The huge difference between the observed value of the cosmological {\it classical} vacuum energy $\Lambda$ {\it today} and the  {\it theoretically} evaluated value of the {\it quantum} particle physics vacuum $\Lambda_Q$, must correctly and physically be like that, because the two values correspond to two huge different physical vacua and eras. The observed $\Lambda$ value today corresponds to the classical, large and dilute (mostly {\it empty}) universe today, (termed voids and supervoids in cosmological observations, termed vacuum space-time in classical gravitation), and this is consistent  with the very low observed  $\Lambda$  vacuum value, ($10^{-122}$ in Planck units), while the computed quantum value $\Lambda_Q$  corresponds to the quantum, small and highly dense energetic universe in its far (trans-Planckian) past, and this is consistent  with its extremely high, trans-Planckian, value ($10^{122}$ in Planck units).
As is well known, the theoretical value $\Lambda_Q\simeq 10^{122}$ is clearly trans-Planckian, this value corresponds and fits correctly the value of $\Lambda_Q$ in the far past trans-Planckian era and its physical properties: quantum size and time $10^{-61}$, quantum  (Gibbons-Hawking) temperature $10^{61}$ and entropy $10^{-122}$. Consistently too, the trans-Planckian era provides the quantum precursor of inflation from which the known classical/semiclassical inflation era, its CMB observables and quantum corrections are recovered in agreement with the set of well established cosmological observations.

\medskip

Starting from quantum theory to reach the Planck scale and trans-Planckian domain (instead of starting from classical gravity by quantizing general relativity) reveals successful with novel results, {\it "quantum relativity"} and quantum 
space-time structure \cite{Sanchez2019},\cite{Sanchez2},\cite{Sanchez3}. Beyond the classical-quantum duality of the space-time, the space-time coordinates can be promoted to quantum non-commuting operators:
comparison to the harmonic oscillator and global phase space
is enlighting, the  
hyperbolic quantum space-time structure generates the {\it quantum light cone}:
The classical space-time null generators $X = \pm T$ {\it dissapear} 
at the quantum level due to the relevant $[X,T]$ conmutator which is {\it always} non-zero,  
a {\it new} quantum vacuum region beyond the Planck scale emerges.

\medskip

{\bf In this paper} we analyze the new vacuum quantum region inside 
the Planck scale hyperbolae which delimitate the quantum light cone.
The effect of the zero point (vacuum) quantum energy bends the space-time
and produces a constant curvature de Sitter background. We find the quantum discrete levels in the
cosmological vacuum trans-Planckian region and in the post-Planckian one.
The quantum light cone is generated by the quantum Planck hyperbolae $X^2 - T^2 = \pm \; [X, T]$ due 
to the quantum uncertainty $\Delta X \Delta T$ or non-zero commutator $[X,T]$, the classical light cone 
generators $X = \pm T$ being a particular case of it. 
This generalizes the classical known space-time
structure and reduces to it in the classical case (zero quantum commutators). 
In higher $D$ space-time dimensions, the quantum non-commuting space and time coordinates $(X,T)$ 
and the transverse commuting spatial coordinates $X_{\bot j}$
generate the quantum two-sheet hyperboloid $ X^2 - T^2 + X_{\bot j}X_{\bot}^j  = \pm 1;
 \;  j = 2, ... (D-2)$.

\medskip

Interestingly enough, the quantum space-time structure turns out to be {\it discretized}
in quantum hyperbolic levels. For times and lengths larger than the Planck time and length $(t_P,l_P)$,  the levels are  $(X_n^2 , T_n^2) =  (2n+1),\; n = 0,1,2...,$ , (in Planck units) , ($X_n, T_n$) and the {\it mass levels} being $\sqrt {(2n +1)}$. 
The discrete allowed levels from the quantum Planck scale $(X_n, T_n) = 1$, $(n = 0)$ and the quantum levels (low $n$) until the quasi-classical and classical ones (intermediate and large $n$), tend asymptotically (very large $n$) to a continuum classical space-time. 
 In the trans-Planckian domain: times and lengths smaller than the Planck scale, the $(X_n, T_n)$ levels are $1/(2n+1)$, the most higher $n$ being the more excited quantum and trans-Planckian ones.

\medskip

For each level $n = 0, 1, 2, ....,$ the two: post and pre (trans) - Planckian phases are covered: In the post-Planckian universe $t_P \equiv t_{planck} < t \leq t_{today} = 10^{61} t_P$ the levels (in Planck units) for the Hubble constant $H_n$, vacuum energy $\Lambda_{n}$, and gravitational (Gibbons-Hawking) entropy $S_n$ are
\be H_{n} = {1}/\sqrt{(2n + 1)}, \; \; \Lambda_{n} = 1/(2n + 1),\;\; S_n = (2n + 1), \; n = 0, 1, 2, .....  \ee
As $n$ increases, radius and mass increase, $H_n$ and $\Lambda_n$ decrease, $S_n$ increases and {\it consistently} the universe {\it classicalizes}.  
In the pre-Planckian (trans-Planckian) phase $10^{-61} t_P \leq t \leq t_P$, the quantum trans-Plankian levels ($Q$ denoting quantum) are: 
\be H_{Qn} = \sqrt{(2n + 1)},\; \; \Lambda_{Qn} = (2n + 1),\;\; S_{Qn} = 1/(2n + 1), \; n = 0, 1, 2, ..... \ee 
The scalar curvature levels in the respective phases being  $R_{Qn} = (2n + 1)$ and $R_{n} = 1 / (2n + 1)$. The $n$-levels cover {\it all} scales from the remote past highly excited trans-Planckian level $n = 10^{122}$ with maximum curvature $R_Q = 10^{122}$,  vacuum $\Lambda_Q = 10^{122}$ and minimum entropy $S_Q =  10^{-122}$, $n$ decreases passing the Planck level $n = 0$: $H_{planck} = 1 = \Lambda_{planck} = S_{planck}$ and enters the post-planckian phase: $n = 1, 2, ...n_{infl} = 10^{12} ,... n_{cmb} = 10^{114},... n_ {reion}= 10^{118}, ...n_{today} = {122}$  with the most classical value $H_{today} = 10^{-61}$, $\Lambda_{today} = 10^{-122}$, $S_{today} =  10^{122}$. 

\medskip

The space-time (the arena of events) in the quantum domain is described by {\it a quantum algebra} of space-time position and momenta: We implement the {\it Snyder-Yang} algebra in the cosmological context thus yielding a consistent group-theory realization of quantum discrete de Sitter space-time, classical-quantum gravity duality and its symmetry with a clarifying unifying picture:
Our complete (classical and quantum) length $L_{QH} (l_P, L_H) = L_Q + L_H = l_P ( L_H / l_P + l_P/ L_H)$, where $L_H$ is the classical universe radius and $ L_Q = l_P^2/L_H$ is its quantum size (the Compton length), turns out to be the appropriate length for the two-parameter Snyder-Yang algebra, thus providing a quantum operator realization of the complete de Sitter universe including the quantum trans-Planckian and classical late de Sitter phases.

\medskip

{\bf This paper is organized as follows:} 
In section II we describe the standard model of the universe extended back in time before inflation, thus covering its different phases: classical, semiclassical and quantum -Planckian and transplanckian- domains and their properties including the gravitational entropy and temperature.
Section III summarizes our arguments clarifying the cosmological constant problem as a vacuum energy and the gravitational entropy precisely covering  the quantum (trans-Planckian and Planckian),  the semiclassical and the classical gravitational regimes. Sections VI, VII, VIII, IX and X include  genuinely new material and the new results of the current manuscript: In Sections IV and V we describe the classical, quantum dual and complete de Sitter universe covering the different de Sitter regimes. Sections VI and VII show the link of de Sitter universe and the cosmological constant to the harmonic oscillator. Section VIII shows the link of the space-time structure to the phase space (classical and quantum) harmonic oscillator and describes the quantum space-time discrete levels. In Sections VIII and IX we find the quantum discrete levels of the universe: size, time, vacuum energy, Hubble constant, entropy and their properties from the very early trans-Planckian phase to today dark energy. In Section X  we describe the Snyder-Yang algebra as a group-theory realization of quantum discrete de Sitter space-time and of classical-quantum gravity duality symmetry. Section XI provides discussion and clarification on the results of this approach, in particular the time discrete levels and those of  the vacuum energy, varying cosmological constant and the cosmic evolution in the context of QFT. Section XII summarizes remarks and conclusions and the clarifying unifying picture we obtained.

\section {The Standard Model of the Universe before Inflation}

The set of robust cosmological data (cosmic microwave background, large scale
structure and deep galaxy surveys, supernovae observations, measurements of the Hubble-Lemaitre 
constant and other data) support the standard (concordance) model of the universe and place de 
Sitter (and quasi-de Sitter) stages as a real part of it \cite{WMAP1},\cite{WMAP2},\cite{Riess},\cite{Perlmutter},\cite{Schmidt}\cite{DES},\cite{Planck6}. Moreover, the physical classical, 
semiclassical and quantum Planckian and trans-Planckian de Sitter regimes are particularly 
important for several reasons:

\medskip

{\bf (i)} The classical, present time accelerated expansion of the Universe and its associated dark energy or
 cosmological constant in the today era: classical cosmological de Sitter regime.

\medskip

{\bf (ii)} The semiclassical early accelerated expansion of the Universe and its associated 
Inflation era: semiclassical cosmological de Sitter (or quasi de Sitter) regime (classical general 
relativity plus quantum field fluctuations.)

\medskip

{\bf (iii)} The quantum, very early stage preceeding the Inflation era: Planckian and super-Planckian 
quantum era. Besides its high conceptual and fundamental physics interest, this era could be of realistic 
cosmological interest for the test of quantum theory itself at such extreme scales, as well as for the 
search of gravitational wave signals from quantum gravity for e-LISA \cite{LISA} for instance, after the 
success of LIGO \cite{LIGO},\cite{DESLIGO}. In addition, this quantum stage should be relevant in 
providing quantum precursors and consistent initial states for the semiclassical (fast-roll and slow 
roll) inflation, and their imprint on the observable primordial fluctuation spectra for instance. 
Moreover, a novel result is that this quantum era allows a clarification of dark 
energy as the vacuum cosmological energy or cosmological constant.

\medskip

{\bf (iv)} de Sitter is a simple and smooth constant curvature vacuum background without any physical 
singularity, it is  maximally symmetric and can be described as a hyperboloid embedded in Minkowski space-
time with one more spatial dimension. Its radius, curvature and equivalent density are described in terms 
of only one physical parameter: the cosmological constant.

\medskip

The lack of a complete theory of quantum gravity (in field and in string theory) 
does not preclude to explore and describe quantum  Planckian and trans-Planckian regimes.
Instead of going from classical gravity to quantum gravity by quantizing general relativity, (is not our aim here to review it), we start from quantum physics and its foundational milestone:
the classical-quantum (wave-particle) duality, and extend it to include gravity and the Planck scale 
domain, namely,  wave-particle-gravity duality,  
(or classical-quantum gravity duality), \cite{Sanchez2019}, \cite{Sanchez2003-1}. As a consequence, the different gravity regimes are covered: 
classical, semiclassical and quantum, together with the Planckian and trans-Planckian domain and the elementary particle mass range as well. This duality is {\it universal}, as the 
wave-particle duality, this does not rely 
on the number of space-time dimensions (compactified or not), nor on any symmetry, isometry nor on any other {\it at priori} condition. It includes the known classical-quantum duality as a special case and allows a general clarification from which physical understanding and cosmological 
results can be extracted. This is not an assumed or conjectured duality. 

\medskip

{\bf The standard model of the universe extended to earlier trans-Planckian eras}. The gravitational history of the universe before the Inflation era and the current picture can be  extended by including the quantum precursor phase within the standard model of the universe in agreement with observations. Quantum physics is more complete than classical physics 
and contains it as a particular case: It adds a new quantum Planckian and
trans-Planckian phase of the Universe from the Planck time $t_P$ until the extreme past $10^{-61} t_P$, which is an upper bound for the origin of the Universe, with energy  $H_Q = 10^{61} h_P$, in a similar manner the present age is a lower bound to the (unknown) future age. 

\medskip

The classical large dilute Universe today and the highly dense very early quantum trans-Planckian Universe are classical-quantum duals of each other in the precise meaning of the classical-quantum duality. This means the following:
The classical Universe today $U_{\Lambda}$ is clearly characterized by the set of physical gravitational 
magnitudes or observables (age or size,  mass, density, temperature, entropy) $\equiv (L_\Lambda, M_\Lambda, \rho_\Lambda, T_\Lambda, S_\Lambda)$:
\begin{equation}\label{ULambda1}
U_{\Lambda} = (L_\Lambda, M_\Lambda, \rho_\Lambda, T_\Lambda, S_\Lambda)
\end{equation}
The highly dense very early quantum Universe $U_Q$ is characterized by the corresponding set of quantum dual physical quantities $(L_Q, M_Q, \rho_Q, T_Q, S_Q)$ in the precise meaning of the classical-quantum duality:
\begin{equation} \label{UQ1}
U_Q = (L_Q, M_Q, \rho_Q, T_Q, S_Q)
\end{equation}
\begin{equation} \label{Udual1}
U_Q =  \frac{u_P^2}{U_\Lambda}, \qquad u_P = (l_P, m_P, \rho_P, t_P, s_P)
\end{equation} 

$u_P$ standing for the corresponding quantities at the fundamental constant Planck scale,
the {\it crossing scale} between the two main (classical and quantum) gravity domains.
The classical $U_{\Lambda}$ and quantum $U_Q$ Universe eras or regimes (classical/semiclassical eras of the known Universe and its quantum Planckian and trans-Planckian very early phases), satisfy Eqs.(\ref{ULambda1})-(\ref{Udual1}). The {\it total} Universe $U_{Q\Lambda}$ is composed by their classical/semiclassical and quantum phases:
\begin{equation} \label{Utotal1} U_{Q\Lambda} = \left(\; U_Q  +  U_{\Lambda} + u_P \;\right)
\end{equation} Subscript $\Lambda$ -or equivalently $H$ for Hubble Lemaitre- stands for the classical magnitudes, $Q$ stands for Quantum, and $P$ for the fundamental Planck scale constant values.

In particular, the quantum dual de Sitter universe $U_Q$ is generated  from the classical de Sitter universe  
$U_{\Lambda}$ through Eqs.(\ref{ULambda1})-(\ref{Utotal1}): {\it classical-quantum de Sitter 
duality}. The {\it total} (classical plus quantum dual) de Sitter 
universe $U_{Q\Lambda}$ endowes automatically a  {\it classical-quantum de Sitter symmetry}. 
This includes in particular the classical, quantum and total de Sitter temperatures and entropies 
and allows to characterize in a complete and  precise way the different 
classical, semiclassical, quantum Planckian and super-Planckian de Sitter regimes. 
$H$ stands for the classical Hubble-Lemaitre constant, or its equivalent $\Lambda = 3 \left ({H}/{c}\right)^2$. $H_Q$ (or $\Lambda_Q$) stands for quantum dual, and $Q\Lambda$ (or $QH$) for the total or complete quantities including the both ones.

The size of the Universe is the gravitational length $L_\Lambda = \sqrt{ 3/\Lambda}$
in the classical regime, it is the quantum Compton length $L_Q$ in the quantum dual regime (which is the full quantum Planckian and super-Planckian regime), and it is the Planck length $l_P$ at the fundamental Planck scale: the {\it crossing scale}. The {\it total}  (or complete) size $L_{Q \Lambda}$ is the sum of the two components. Similarly, the horizon acceleration (surface gravity) $K_{\Lambda}$ of the universe in its classical gravity regime becomes the quantum acceleration $K_Q$ in the quantum dual gravity regime. The temperature $T_{\Lambda}$, measure of the classical gravitational length or mass becomes the quantum temperature $T_Q$ (measure of the quantum size or compton length) in the quantum  regime. Consistently, the  Gibbons-Hawking temperature  
is {\it precisely} the quantum temperature $T_Q$. Similarly,  the classical/semiclassical gravitational area or entropy $S_\Lambda$ (Gibbons-Hawking entropy) has its quantum dual $S_Q$ in the quantum gravity (Planckian and trans-Planckian) regime. In sections III and VIII we discuss the concept of gravitational entropy and its expressions in the different gravity regimes.
The concept of gravitational entropy is {\it the same} for any of the gravity regimes: $ Area / 4 l_P^2$ in units of $k_B$.
For a classical object of size $L_\Lambda$, this is the classical area $A_\Lambda = 4 \pi L_\Lambda^2 $. For a quantum object of quantum size $L_Q$, this is the area $A_Q = 4 \pi L_Q^2 $:
\begin{equation} \label{AHL}
A_\Lambda = a_P\left(\frac{L_\Lambda}{\lambda_P}\right)^2,
\qquad
A_Q = a_P\left(\frac{\lambda_P}{L_\Lambda}\right)^2 
 = \frac{a_P^2}{A_\Lambda},\qquad  a_P = 4 \pi\; l_P^2 
\end{equation}
$a_P$ being the Planck area. The corresponding gravitational entropies $S_\Lambda$, $S_Q$ are
\begin{equation} \label{SH}
S_\Lambda = \frac{\kappa_B}{4} \;\frac{A_\Lambda}{l_P^2}, \qquad S_Q = \frac{\kappa_B}{4}\;\frac{A_Q}{l_P^2} 
\end{equation}
And the total (classical and quantum) gravitational entropy $S_{Q\Lambda}$ being given by
\begin{equation} \label{SQH}
S_{Q\Lambda} =  2\;[\; s_P + \frac{1}{2}\;( S_\Lambda + S_Q )\;], 
\qquad s_P = \frac{\kappa_B}{4} \frac{a_P}{l_P^2} = \pi \kappa_B,
\end{equation}
$s_P$ being the Planck entropy.

\section{Classical, Semiclassical and Quantum Vacuum Energy of the Universe}

The classical universe today $U_\Lambda$ is precisely  a {\it classical dilute gravity vacuum dominated by voids and supervoids} as shown by observations \cite{VoidsHistory},
\cite{Voids1}, \cite{VoidsPRL} whose observed $\rho_\Lambda$ or $\Lambda$ value today \cite{Riess},\cite{Perlmutter},\cite{Schmidt},\cite{DES},\cite{Planck6}
is {\it precisely} 
the classical dual of its quantum precursor values $\rho_Q,\Lambda_Q$ in the quantum very early precursor vacuum $U_Q$ as determined by Eqs.(\ref{ULambda1})-(\ref{UQ1}). The high density $\rho_Q$ and cosmological constant $\Lambda_Q$ are 
precisely the quantum particle physics trans-Planckian value $10^{122}$. This is precisely expressed by Eqs.(\ref{ULambda1})-(\ref{UQ1}) applied to this case: 
\begin{equation} \label{LambdaHvalue1} 
\Lambda = 3 H^2 = \lambda_P  \left (\frac{H}{h_P}\right)^2 = \lambda_P \left (\frac{l_P}{L_H}\right)^2
= (2.846 \pm 0.076) \; 10^{-122}\; m_P^2
\end{equation}
\begin{equation} \label{LambdaQvalue1} 
\Lambda_Q = 3 H_Q^2 = \lambda_P \left (\frac{h_P}{H}\right)^2 = \lambda_P \left (\frac{L_H}{l_P}\right)^2
= (0.3516 \pm 0.094) \; 10^{122}\;h_P^2 
\end{equation}
\begin{equation} \label{LambdaQLambda1} 
\Lambda_Q = \frac{\lambda_P^2}{\Lambda}, \qquad \lambda_P = 3 h_P^2
\end{equation}
The quantum dual value $\Lambda_Q$ is {\it precisely} the quantum vacuum value $\rho_Q = 10^{122}\; \rho_P$ obtained from particle physics:
\begin{equation} \label{rhoQii1}
\rho_Q = \rho_P \left(\frac{\Lambda_Q}{\lambda_P}\right) 
= \frac{\rho_P^2}{\rho_\Lambda} = 10^{122}\; \rho_P
\end{equation}

 In the last r.h.s. of Eqs.(\ref{LambdaHvalue1})-(\ref{LambdaQLambda1}) the the data refs \cite{Riess},\cite{Perlmutter},\cite{Schmidt},\cite{DES},\cite{Planck6} have been used,  which we also {\it link to the gravitational entropy and temperature of the universe}. 
The  {\it complete} total vacuum energy density 
$\rho_{Q\Lambda}$ or $\Lambda_{Q\Lambda}$ is the sum of its classical and quantum components (corresponding to the classical today era and its quantum  Planckian and trans-Planckian precursor):
\begin{equation} \label{LambdaQLambda3}
\Lambda_{Q \Lambda} = \lambda_P \left (\;    
  \frac{\Lambda}{\lambda_P} + \frac{\lambda_P}{\Lambda} + 1\;\right) = 
\lambda_P \;(\; 10^{-122} + 10^{+122} + 1\;)
\end{equation}

The observed $\Lambda$ or $\rho_\Lambda$ today is the {\it classical gravity vacuum} value 
in the classical universe $ U_\Lambda$ {\it today}. Such observed value must be 
consistently in such way because of the {\it large classical} size of the universe today  $L_
\Lambda = \sqrt{3/\Lambda}$, and of the empty or vacuum  dilute state today dominated by {\it 
voids and supervoids} as shown by the set of large structure observations \cite{VoidsHistory},
\cite{Voids1}, \cite{VoidsPRL}. This is one main physical reason for such a {\it low} $\Lambda$ 
value at the present age today  $10^{61} t_P$. Its precursor value and 
density $\Lambda_Q , \rho_Q$ is a high super-Planckian value precisely because this is a high density 
very early {\it quantum cosmological vacuum} in the extreme past $10^{-61} t_P$ of the quantum trans-Planckian precursor phase $U_Q$.

\medskip

The quantum vacuum density $\Lambda_Q = \rho_Q = 10^{122}$ 
(in Planck units) in the precursor trans-Planckian phase $U_Q$ at $10^{-61} t_P$, 
(the extreme past), became  the classical vacuum density $\Lambda = \rho_\Lambda = 10^{-122}$ 
in the classical universe $U_\Lambda$ today at $10^{61} t_P$. 
The trans-Planckian value is consistently in such way because is a extreme quantum gravity (trans-Planckian) vacuum 
in the extreme quantum past $10^{-61} t_P$ with minimal entropy $S_Q = 10^{-122} = \Lambda = \rho_\Lambda$.
Eqs.(\ref{LambdaHvalue1}) to (\ref{rhoQii1}),(\ref{LambdaQLambda3}) concisely {\it explain  why} 
the classical gravitational vacuum $\Lambda$ or $\rho_\Lambda$ {\it coincides} with such observed {\it low value} $10^{-122}$ in Planck units, and {\it why} their corresponding quantum gravity precursor vacuum has such extremely {\it high} trans-Planckian {\it value} $10^{122}$. The classical gravitational entropy $S_\Lambda$ today has {\it precisely} such high value:  
\begin{equation} \label{SrhoQLambda}
S_{\Lambda} = s_P \;\left(\frac{\rho_Q}{\rho_P}\right) = s_P \;\left(\frac{\lambda_P}{\Lambda}\right) = s_P \; 10^{+122}
\end{equation}
\begin{equation} \label{SQrhoLambda}
S_Q = s_P \; \left(\frac{\rho_\Lambda}{\rho_P}\right) = s_P \; \left(\frac{\Lambda}{\lambda_P}\right) = s_P \; 10^{-122}
\end{equation} 
The {\it total} $Q\Lambda$ (classical and quantum) gravitational entropy $S_{Q\Lambda}$ derives from the general expression 
$$S_{Q\Lambda} =  (A_{Q\Lambda} / 4 l_{l_P}^2) k_B, $$
where the total area  
$$ A_{Q\Lambda} =  4 \pi L_{Q\Lambda}^2  = 4 \pi (L_Q + L_\Lambda)^2, $$  
expresses as $A_{Q\Lambda} = A_Q  +  A_\Lambda + 2 a_P$.          
Recall that  $L_Q = l_P^2 / L_\Lambda$   and  $a_P  =  4\pi l_P^2$.

As a consequence: 
\begin{equation} \label{Stotalvalue}
S_{Q\Lambda} =  2\;s_P + S_{\Lambda} + S_Q 
= 2 \;s_P \;[\; 1 + \frac{1}{2}(10^{+122} + 10^{-122})\;],
\end{equation}
$s_P$ being the Planck entropy. The {\it total} $Q\Lambda$ gravitational entropy turns out the sum of the three components as it must be: classical (subscript $\Lambda$), quantum (subcript $Q$) and Planck value (subscript $P$) corresponding 
to the tree gravity regimes. The term $2s_P$ arises from the duality between the quantum and classical lengths $L_Q$ and $L_\Lambda$ across the Planck scale. The factor $2$ reflects the complete $Q\Lambda$ covering: the Planck scale being the bordering or crossing scale common to the two (classical and quantum) $Q$ and $\Lambda$ domains. 

\medskip

The gravitational entropy $S_{\Lambda}$ of the present time large {\it classical universe} is a very {\it huge number}, 
consistent with the fact that the universe today contains a very huge amount of information. Moreover, to reach such a huge size and entropy today $10^{+122}$, the universe in its very beginning should have been in a hugely energetic initial vacuum $10^{+122}$.

\medskip

{\bf A whole picture.} Overall, a consistent 
unifying picture of the gravitational cosmic history
through  {\it its vacuum energy} does emerge from 
the extreme past quantum trans-Planckian, Planckian 
and post-Planckian phases: semiclassical (inflation)
and classical today phases and their relevant
physical magnitudes: size, age, gravitational
entropy and temperature, all in terms of the vacuum
energy. This sheds light on inflation and dark
energy. The whole duration (of the trans-Planckian
plus post-Planckians eras) is precisely
$10^{-61}\; t \leq t \leq 10^{+61}$ (in Planck units
$t_P = 10^{-44}$ sec). That is to say, 
{\it each} time component {\it naturally} dominates in
each phase: classical time 
component $ 10^{+61}$ in the classical era, quantum
Planck time $t_P$ 
in the quantum preceding era. The present time of
the universe at $10^{+61} t_P$, is {\it a lower
bound} for the future (if any) age 
of the universe, the remote past quantum precursor
equal to $10^{-61}\; t_P$, is an {\it upper bound} 
for the origin of the universe. The known classical/semi-classical inflation era which occurred at about $10^
{+6} t_P$, $H = 10^{-6} h_P$ has a preceding quantum era at $10^{-6} t_P$,  $H = 10^{6}h_P$ which is 
in fact a semi-quantum era ('low $H$' with respect to the extreme past trans-Planckian state $H = 10^{61} h_P$), and 
similarly, for any of the other known eras in the classical post-Planckian universe: they have a corresponding quantum precursor era in the trans-Planckian phase. This appears to be the way in which the universe has evolved.\\
The {\bf total or complete (classical plus quantum)} physical quantities are invariant under the classical-quantum duality: 
$H \leftrightarrow Q$ (or $\Lambda \leftrightarrow Q$) as it must be: This means physically that: (i) what occurred in the quantum phase 
before $t_P$ {\it determines} through  
Eqs.(\ref{ULambda1})-(\ref{Utotal1}) what occurred in the classical 
phase after $t_P$. And: (ii) what occurred in the quantum phase before the planck time $t_P$ is the {\it same observable} which occurred after 
$t_P$ but in a different physical state in the precise meaning of Eqs.(\ref{ULambda1})-(\ref{Utotal1}). 
That is to say: The quantum quantities in the phase before 
$t_P$, are the {\it quantum precursors} of the classical/semiclassical quantities after $t_P$. 
As the wave-particle duality at the basis of quantum mechanics, the 
wave-particle-gravity 
duality is reflected in all cosmological eras and its associated 
quantites, temperatures and 
entropies. Cosmological evolution goes from a quantum trans-Planckian 
vacuum energy 
phase to a semiclassical accelerated era (de Sitter inflation), then 
to the classical known
eras until the present classical de Sitter phase. The 
classical-quantum or wave-particle-gravity duality specifically 
manifests in this evolution, between the different gravity regimes, 
and could be view 
as a mapping between asymptotic (in and out) states characterized by 
sets $U_Q$ and $U_\Lambda$ and thus as a Scattering-matrix 
description.

\section{Classical and Quantum Dual de Sitter Universes}

De Sitter space-time in $D$ space-time dimensions is 
the hyperboloid embedded in $(D+1)$ dimensional Minkowski space-time:
\begin{equation} \label{dS}
X^2 - T^2 + X_j X^j + Z^2 = L_H^2, \qquad j= 2, 3, ...(D-2)
\end{equation}
$L_H$ is the classical radius or characteristic length of the de Sitter universe. 
The scalar curvature $ R$ is constant. Classically: 
$$
L_H = {c}/{H}, \qquad R =  H^2 D(D-1) = \frac{2D}{(D-2)}\;\Lambda, 
\qquad \Lambda = \frac{H^2}{2} (D-1) (D-2)
$$

\medskip

A mass $M_H$ can be associated to $L_H$ or $H$, such that (D = 4 for simplicity):
\begin{equation} \label{LH}
L_H = {G M_H} / {c^2} \equiv L_G, \;\;\qquad M_H = {c^3} / {(GH)} 
\end{equation}
The corresponding quantum dual magnitudes $L_Q$, $M_Q$ are:
\begin{equation} \label{MH}
L_Q = \frac{\hbar}{M_H c} = \frac{\hbar\; G H}{c^3} = \frac {l_P^2}{L_H}\;, \;\;\qquad M_Q = \frac{\hbar H}{c^2} = \frac{m_P^2}{M_H}
\end{equation}
\begin{equation} \label{MQ}
{\text ie,} \quad  L_Q = \frac{l_P^2}{L_H}\;, \;\;  \qquad M_Q = \frac{m_P^2}{M_H}
\end{equation}
 $l_P$ and $m_P$ being the Planck length and Planck mass respectively:
\begin{equation}\label{lp}
l_P = \sqrt{{\hbar\;G} /{c^3}}\;, \;\; \qquad m_P = \sqrt{{c\;\hbar}/{G}}
\end{equation}

The quantum dual Hubble constant  $H_Q$ and the quantum curvature $R_Q$ are:
\begin{equation}\label{dualQH}
H_Q = {h_P^2} / {H} , \qquad R_Q = {r_P^2} / {R}, 
\qquad \Lambda_Q = {\lambda_P^2} / {\Lambda}
\end{equation}  

where $ h_P, r_P, \lambda_P $ are the Planck scale values of the Hubble constant, 
scalar curvature and cosmological constant respectively:
\begin{equation}\label{P}
h_P = {c} / {l_P} ,\qquad r_P = h_P^2 \;D (D-1) 
, \qquad \lambda_P = \frac{h_P^2}{2} \;(D-1) (D-2) 
\end{equation}  
\begin{equation}\label{PD4}
h_P = c^2 \sqrt{{c}/{\hbar G}},\qquad r_P =  12 \; h_P^ 2 = 4 \;\lambda_P ,\qquad 
\lambda_P = 3 \; \left({c^5} / {\hbar G}\right),  \qquad (D=4)
\end{equation} 

\section {Total de Sitter Universe and its Duality Symmetry}

The classical and quantum lengths: $L_H$, $ L_Q$ can be extended 
to a more complete length $L_{QH}$ which contains both: the Q and H lengths):
\begin{equation}\label{LQH}
L_{QH} =  (L_H + L_Q) = l_P \; (\frac{L_H}{l_P} + \frac{l_P}{L_H}) 
\end{equation}
and we have then : 
\begin{equation}\label{ZQH}
X^2 - T^2 + X_j X^j + Z^2 = L_{QH}^2 = 
 2\;l_P^2 \left[1 + \frac{1}{2}\; [\;(\frac{L_H}{l_P})^2 + (\frac{l_P}{L_H})^2\;]\right]
\end{equation}
with   $j = 2,3, ...(D-2)$. $Z$ is the extra coordinate for the embedding of de Sitter space-time in Minkowski space-time.  

\medskip

Eq.(\ref{ZQH}) quantum generalize de Sitter space-time including the classical, 
semiclassical and quantum Planckian and trans-Planckian de Sitter regimes  as well. 
It contains two non-zero lengths $(L_H, L_Q)$ or two relevant scales ($H$,  $l_P$) 
enlarging the possibilities for the space-time phases, thus:

\medskip

\begin{itemize}
\item{For $ L_H >> l_P$, ie $L_Q << L_H$, Eq.(\ref {ZQH}) yields the classical de Sitter space-time. 
For intermediate $L_H$ values between $l_P$ and $L_Q $ it yields
the semiclassical de Sitter space-time. }

\item{For $ L_H = l_P $ ie $L_Q =  l_P = L_{QH}$, Eq.(\ref{ZQH}) yields
 the planck scale de Sitter hyperboloid.}
\item{For $ L_H << l_P $, ie $L_Q >> L_H $ it yields the highly quantum de Sitter regime, deep 
inside the planck domain.}
\end{itemize}

\medskip

$H = c / L_H $ is ($ c ^{-1} )$ times the surface gravity (or gravity acceleration) 
of the classical de Sitter space-time. Similarly, $ H_Q = c / L_Q $ 
and $H_{QH} = c / L_{QH}$ are the surface gravity in the quantum and whole  QH de Sitter phases respectively.
Similarly, Eq. (\ref{LQH}) and Eqs (\ref{LH})-(\ref{MQ}), yield for the mass:
\begin{equation}\label{MQH}
M_{QH} =  (\;M_H + M_Q\;) = m_P \;(\;\frac{M_H}{m_P} + \frac{m_P}{M_H}\;)
\end{equation}
\begin{equation}\label{MLQH}
\frac{M_{QH}} {m_P} = m_P \;(\;\frac{L_H}{l_P} + \frac{l_P}{L_H}\;) = \frac{L_{QH}}{l_P}
\end{equation}

$M_{QH}/m_P$ and  $L_{QH}/l_P$ both have the same expression with respect to their respective Planck values.

\bigskip

{\bf The complete QH Hubble constant $H_{QH}$, curvature $R_{QH}$ and $\Lambda_{QH}$.} 

\bigskip

The total (classical and quantum) QH Hubble constant $H_{QH}$, curvature $R_{QH}$ and $\Lambda_{QH}$ 
follow from the QH de Sitter length $L_{QH}$ Eq.(\ref{LQH}):
\begin{equation}       \label{QH1}
H_{QH} = \frac{c}{L_{QH}}, \qquad  R_{QH} = H_{QH}^2\; D\;(D-1), 
\qquad  \Lambda_{QH} =  \frac{H_{QH}^2}{2} \;(D-1) (D-2)
\end{equation}
where from Eqs.(\ref{LQH}) and (\ref{dualQH}):
\begin{equation}    \label{QH2}
H_{QH} = \frac{H}{ [\;1 + (l_P H / c)^2\;]}, \qquad H_{QH}/ h_P = \frac{(H/h_P)}{ [\;1 + (H / h_P)^2\;]},
\qquad h_P = c / l_P
\end{equation}

\medskip

which exhibit the {\it symmetry} of $H_{QH}/ h_P$ under $(H/h_P) \rightarrow (h_P/H)$, 
ie under $ H \rightarrow H_Q = (h_P^2/H)$ :
\begin{equation}    \label{symQH}
H_{QH} (H / h_P) = H_{QH} (h_P / H)
\end{equation}

The classical $H$ and quantum $H_Q$ are classical-quantum duals of each other through the Planck scale $h_P$, but the total $H_{QH}$ is
{\it invariant}. And similarly, for the total quantum curvature  $R_{QH}$ and cosmological constant $\Lambda_{QH}$ Eq.(\ref{QH1}):
\begin{equation}\label{symRQH}
R_{QH}(H/h_P) = R_{QH}(h_P/H),  \qquad  \Lambda_{QH}(H/h_P) = \Lambda_{QH}(h_P/H)
\end{equation}
where:
\begin{equation}\label{R1}
R_{QH} = \frac{R_H}{ [\;1 + R_H /r_P \;]^2}  =  \frac{R_Q}{[\;1 + R_Q /r_P \;]^2},\qquad r_P = 12\;h_P^2
\end{equation}

\begin{equation}\label{L1}
\Lambda_{QH} = \frac{\Lambda_H}{[\;1 + \Lambda_H /\lambda_P \;]^2}  =  
 \frac{\Lambda_Q}{[\;1 + \Lambda_Q /\lambda_P \;]^2},\qquad \lambda_P = 3\;h_P^2
\end{equation}

\bigskip

The classical $ H/h_P << 1$, quantum $ H/ h_P >> 1$ and Planck $H/h_P = 1$ 
regimes are clearly exhibited in the  QH expressions Eqs (\ref{QH1}), Eq.(\ref{QH2}):
\begin{equation}\label{sHP}
H_{QH \;(H << h_P)} =   H \;[\; 1 - (H / h_P)^2 \;] + O\;(H / h_P)^4  
\end{equation}
\begin{equation}\label{HP}
H_{QH} \; (H = h_P)  = \frac{h_P}{2}, \; \; h_P= c/l_P
\end{equation}
\begin{equation}\label{bHP}
H_{QH \; (H >> h_P)} =  (h_P^2/H)\; [ 1 - (h_P/H)^2 ] \;+\; O (h_P/H)^4
\end{equation}

The three above equations show respectively the three different de Sitter phases:
\begin {itemize}
\item {The classical gravity de Sitter universe (with lower curvature  than the Planck scale $r_P
$) {\it outside}
the Planck domain $(l_P < L_H < \infty)$.} 
\item {The Planck curvature de Sitter state  $(R_H = r_P, \;\; L_ H = l_P)$}
\item {The highly quantum or high curvature
($R_H >> r_P$) de Sitter phase {\it inside} the quantum gravity Planck domain ($0 < L_H \leq  
l_P$).}
\end {itemize}

Is natural here to define the dimensionless magnitudes: 
\begin{equation}\label{dimless}
{\cal L} \equiv L_{QH}/l_P, \qquad {\cal M}\equiv 
M_{QH}/m_P, \qquad  {\cal H} \equiv  H_{QG}/h_P, 
\qquad l\equiv L_H /l_P, \qquad  h \equiv H / 
h_P = l^{-1}
\end{equation} 
in terms of which,  Eqs (\ref{LQH}),(\ref{MQH}) and (\ref{QH2}) 
and their duality symmetry Eqs (\ref{symQH}), 
(\ref{symRQH}) simply read:
\begin{equation}\label{dimless2}
{\cal L} =  (l + \frac{1}{l}) = {\cal M},\qquad 
{\cal H} = \frac{1}{(l + \frac{1}{l})} = {\cal L}^{-1}
\end{equation} 
\begin{equation}\label{sym}
{\cal L} (l^{-1}) ={\cal L} (l), \qquad  {\cal M} (l^{-1}) = {\cal M} (l) 
\end{equation} 
\begin{equation}\label{sym2}
{\cal H} (l^{-1}) ={\cal H} (l), \qquad  {\cal R} (l^{-1}) = {\cal R} (l),
 \qquad  {\mathbf \Lambda} (l^{-1}) = {\mathbf \Lambda} (l)
\end{equation} 

\medskip

The QH magnitudes are complete variables covering both classical and quantum, Planckian and trans-Planckian, domains. 
Similarly, for the classical, quantum and QH de Sitter densities ($\rho_H$, $\rho_Q$, $\rho_{QH}$), $\rho_P$ 
being the Planck density scale):
\begin{equation}\label{rhoH}
\rho_H = \rho_P \left({H}/{h_P}\right)^2 = \rho_P \left({\Lambda}/{\lambda_P}\right), \qquad 
\rho_P = {3 \; h_P^2}/{8 \pi G}, \quad \lambda_P = {3\; h_P^2} /{c^4}
\end{equation} 
\begin{equation}\label{rhoQ}
\rho_{Q} = \rho_P \left({H_Q}/{h_P}\right)^2 = \rho_P  \left({\Lambda_Q}/{\lambda_P}\right) = {\rho_P^2}/{\rho_H}
= \rho_P \left({h_P}/{H}\right)^2 = \rho_P \left({\lambda_P}/{\Lambda}\right)
\end{equation}                             
\begin{equation}\label{rhoQH}
\rho_{HQ} = \rho_{H} + \rho_{Q} = \rho_P \left({H_{HQ}}/{h_P}\right)^2 = \rho_P \left({\Lambda_{HQ}}/{\lambda_P}\right) 
\end{equation}
From which it follows that:\begin{equation}
\rho_{HQ}= \frac{\rho_H}{[\; 1 + \rho_H / \rho_P\; ]^2} 
= \frac{\rho_Q}{[\; 1 + \rho_Q / \rho_P \; ]^2}, 
\end{equation}
which satisfies 
$$
\rho_{HQ}\; (\rho_H)  =  \rho_{HQ}\; (\rho_Q) = \rho_{HQ} \;(\rho_P^2 / \rho_H),
$$
For small and high densities with respect to the Planck density $\rho_P$, the QH density $\rho_{QH}$ behaves: 
\begin{equation} \label{RhoQH1} \rho_{QH} \;(\rho_H << \rho_P) =  \rho_H \;[\; 1 - 2 (\rho_H / \rho_P) \;] 
+ O\;(\rho_H / \rho_P)^2  
\end{equation}
\begin{equation} \label{RhoQHP}
\rho_{QH} \; (\rho_H = \rho_Q = \rho_P)  = \frac{1}{4}\rho_P : \;\;\mbox {(Planck scale density)}
\end{equation}
\begin{equation}  \label{RhoQH2}
\rho_{QH} \;(\rho_H >> \rho_P) =   \rho_Q \;[\; 1 - 2 (\rho_Q / \rho_P) \;] + O\;(\rho_Q / \rho_P)^2, 
\end{equation}

corresponding to the classical/semiclassical de Sitter regime (and its quantum corrections), Planck scale de Sitter state 
and highly quantum trans-Planckian de Sitter density. The complete QH de Sitter magnitudes $(L_{QH}, H_{QH}$, $M_{QH})$, 
[and their constant Planck scale values $(l_P, h_P, m_P)$ only depending on $(c, \hbar, G)]$, allow to characterize in a precise
way the classical, semiclassical, Planckian and quantum (super-Planckian) de Sitter regimes:

\begin{itemize}
\item{$L_{QH} = L_{QH}(L_H, L_Q) \equiv L_{QH} (H, \hbar) $ 
yields the {\it whole} (classical/semiclassical, Planck scale and quantum (super-Planckian) 
de Sitter universe.}
\item{$L_{QH} = L_H = L_Q$ yields the Planckian de Sitter state, 
(Planck length de Sitter radius, Planckian vacuum density and Planckian scalar curvature):
$L_H = l_P,  \; H = h_P, \; \lambda_P = 3\; h_P^2, \; R = r_P = 4 \;\lambda_P, 
\quad l_P = \sqrt{(\hbar G / c^3)}$}
\item{$L_{QH} = L_H >> L_Q $,  ie $ L_H >> l_P $ , $ H << h_P $,
 yields the classical de Sitter space-time.}
\item{$L_{QH} = L_Q  >> L_H $, ie $ L_H << l_P $,  $H >> h_P$, 
(high curvature $R >> r_P = 4 \Lambda_P$), \\ 
yields a full quantum gravity trans-Planckian de Sitter phase 
(inside the Planck domain $0 < L_H \leq l_P$).}
\item{$L_{QH} >> L_Q$ ie $L_{QH} \rightarrow \infty$ for $ L_H \rightarrow \infty $, ie 
$H \rightarrow 0 $ ie $\Lambda \rightarrow 0$, 
(zero curvature) yields consistently the classical Minkowski space-time, equivalent to the limit  
$L_Q \rightarrow 0 $ ie $ l_P \rightarrow 0$  ($\hbar \rightarrow 0$).}
\end{itemize}
The three de Sitter regimes are characterized in a complete and precise way:
\begin{itemize}
\item{(i) {\it classical and semiclassical de Sitter regimes}:
(inflation and more generally the whole known 
-classical and semiclassical- universe is within this regime):\\
$ l_p < L_H < \infty$, \; ie $0 < L_Q  < l_P$, $\;$  $0 < H < h_P$,$\;$ $ m_P < M_H < \infty$.} 
\item{ (ii) {\it Planck scale de Sitter state with Planck curvature and Planck radius:} \\
$L_H = l_P , \;\; L_Q = l_P,\;\; H = h_P = c/l_P,\;\; M_H = m_P$.}
\item{(iii)  {\it quantum Planckian and trans-Planckian de Sitter regimes:}
$ 0 < L_H \leq l_P $,\\
 ie $l_P \leq L_Q < \infty$, \;\;$ h_P \leq H < \infty $,\;\; $ 0 < M_H < m_P$.} 
\end{itemize}

\section{de Sitter Universe and the Harmonic Oscillator}

As is known, the Einstein Equations in the presence of a constant vacuum energy (cosmological constant) are
\begin{equation}
  G_{\mu \nu} + \Lambda g_{\mu \nu} = 8\pi G T_{\mu \nu} \; ,
\end{equation}
and the energy-momentum tensor corresponding to the vaccum energy density $\rho$
and pressure $p$ is
\begin{equation}
T_{\mu \nu} =  p g_{\mu \nu } = -\rho g_{\mu \nu }, \quad \;  (p = w \rho, 
\;\; w\equiv -1)
\end{equation}
the vacuum energy being equivalent to a cosmological constant:
$\rho_{\Lambda} = \Lambda c^4 / (8\pi G) $.

As known, de Sitter space-time has {\it constant} scalar space-time curvature:
$$
R = 12\; H^2 = 4\; \Lambda, \;  \qquad \Lambda = 3\; H^2, \qquad \; (D= 4)
$$

We restrict to $ D= 4$. Recall the energy-momentum tensor for massive particles of density $\rho$ plus vacuum constant energy 
(or cosmological constant) $\Lambda$ is:
\begin{equation}\label{tmunu}
T^{\mu}_{\nu} = \rho_{Lambda} \; \delta^{\mu}_{\nu} + \rho \;
\delta^{\mu 0} \; \delta_{\nu 0} \; ,
\quad T \equiv T^{\mu}_{\mu} = 4 \, \rho_{\Lambda} +  \rho  
\end{equation}
where we neglected the pressure to better illustrate our purpose.
The corresponding Einstein equations are
\begin{equation} \label{eins}
R^{\mu}_{\nu} = 8\pi \, G \left(T^{\mu}_{\nu} -
\frac{\delta^{\mu}_{\nu}}2 \; T \right)\; ,\qquad 0 \leq \mu, \nu \leq 3
\end{equation}
and  for non relativistic matter its pressure is neglected with respect to its rest mass. 

In the weak field limit: 
$$g_{00} = 1 + 2 \; V \quad , \quad g_{ik} = 
- \delta_{ik} \; \quad R_0^0 = \nabla^2 V \;,$$
$V$ being the gravitational potential, Einstein's Eqs.(\ref{eins}) become 
\begin{equation} \label{poiL}
\nabla^2 V = 4\pi \, G \, \rho -  8\pi \, G \, \rho_{\Lambda} \; 
\end {equation}
\begin{equation} \label{potentieltot}
V({\vec X}) =  V_{\rho}(X) - \frac{4\pi \, G \,  \rho_{\Lambda}}{3} \, X^2 \; ,
\end{equation}
For a distribution of rest particles of mass $m$, 
$\rho({\vec X}) = m \sum_i \; \delta({\vec X}-{\vec X}_i)$, 
the gravitational potential $V({\vec X})$, gravitational field ${\cal G}$ and potential energy ${\cal U} $ of the system are:
$$
V({\vec X}) =  V(\vec X)_m -\frac{4\pi \, G \, \rho_{\Lambda}}{3} \, X^2, \; 
 \qquad    V(\vec X)_\rho \equiv V(\vec X)_m =  -G \sum_i \frac{m}{|{\vec X}-{\vec X}_i|}.
 $$
\begin{equation} \label{campog}
{\vec {\cal G} ({\vec X}) } = - \nabla V({\vec X}) = {\vec {\cal G}_m } + \frac{8\pi \, G \,
 \rho_{\Lambda}}{3} \; {\vec X} 
\end{equation}
\begin{equation} \label{enerG}
{\cal U} = {\cal U}_m
- \frac{4\pi \, G \, \rho_{\Lambda}}{3} \, m \; \sum_i X_i^2 
\end{equation}
Therefore, the Hamiltonian is equal to:
\begin{equation} \label{hamil}
 \frac{P_i P^i}{\, m^2} + {\cal U} \; = \; \frac{P_i P^i}{\, m^2} - \frac{4\pi\, G \,\rho_{\Lambda}}{3}\, m \,\; X_i^2   
\end{equation}

\medskip

For a relativistic perfect fluid with $T^{\mu \nu} = (p + \rho) u^\mu u^\nu + p g^{\mu \nu}$, and continuity equation $D_{\nu} T^{\mu \nu} = 0$, the $00$ component of the Einstein equations yields a relativistic Poisson equation similar to Eq. (\ref{poiL}) sourced with the addition of the fluid pressure $p$ to the density and to the $\Lambda$ term which remains unchanged. In this case, the potential is coupled to the Euler fluid equations which linearized perturbations for each component can be reduced to an equation of the form $\ddot{\delta} + f \delta  = 0$,  $f$ depending of the unperturbed background fluid components and on the $\Lambda$ term which has always opposite sign to the other component terms. Our interest in this paper not being in the structure formation and evolution but in the vacuum $\Lambda$ energy we will not discuss more here on this case.

\medskip

The cosmological constant energy contribution to the potential energy $\cal U$ Eq.(\ref{hamil}) decreases for increasing values of the particle distances  
$ r_i $ to the center of mass. The gravitational effect of the vacuum zero point energy or cosmological constant push particles outwards 
and equivalently, the last term of the gravitational field Eq.(\ref{campog}) points outward (the repulsive cosmological constant effect). 
The Hamiltonian Eq.(\ref{hamil}) is like that of a harmonic oscillator for a particle of mass $m$ and oscillator constant $ \omega^2 m $. We analyze it in Section VII below. 

\medskip

The non-relativistic particle motion and the relativistic geodesics both exhibit the same runaway behaviour. The non-relativistic approximation reflects well the relativistic particle motion in the de Sitter space-time and its connection to the harmonic oscillator. In the relativistic situation, $g_{00}$ determined by the Einstein Equations for the de Sitter metric entails the harmonic oscillator potential, e.g.
$$d/dr (r g_{00}) = 1 -  (8\pi G\rho_{00} + \Lambda)r^2$$
Parametrization of the de Sitter hyperboloid in terms of the coordinates  $(t, r, \theta, \phi)$: 
$$ T =  H^{-1} (1 - H^2r^2)^{1/2} \;\sinh{ Ht};\;\;\; 
X =  H^{-1} (1 - H^2r^2)^{1/2}\; \cosh {Ht} $$
$$ X_2 = r \cos{\theta};\;\;\; X_3 = r \sin {\phi}\; \cos {\theta};\;\;\;  Z =  r \sin {\phi},$$
yields :$$ ds^2 = - (1 - H^2 r^2)\; dt^2 + (1 - H^2r^2)^{-1} dr^2 + r^2 d\Omega^2$$
and  $X^2 - T^2 = H^{-2} (1 - H^2 r^2)$, containing the (inverted) harmonic oscillator potential. 

\medskip

Our purpose in this paper is to show within a minimal setting the essential features relating de Sitter space-time e.g the cosmological constant or vacuum energy density to the harmonic oscillator.  We are interested in the vacuum energy density and so we do not include all other particle interactions. Self-gravitation interaction among the particles is described by the 
parameter  $(G m^2 N)$ while the interaction with $\Lambda$ is through the parameter $(\Lambda m)$. Their quotient,  namely $\eta = $(vacuum energy / mass) $ =  \Lambda /(m G N)$, $N$ being the number of particles, determines the condition on 
whether one dominates over the other, and clearly our interest in this paper is in the regimes  where the vacuum energy $\Lambda$ dominates over the self-particle interactions, i.e  $\eta \leq 1$, which is the 
condition for interactions be neglected. Of 
course, virialization occurs too for larger $\eta$
for self-gravitating  particles in the presence of $\Lambda$.

\medskip

 In a QFT description, particle production from 
 the vacuum or inflation driven by $\Lambda$ are within this situation of $\Lambda$ we consider, in the pre-Planckian and in the post-Planckian eras. Particle interactions at the Planck scale considered mainly in the context of particle 
 physics,  perturbatively and non-perturbatively, 
 as string  collisions or as point particle QFT 
\cite{tHooft1987}, \cite{AmatiCiafaVenezia}, 
\cite{deVSNPB1989}, \cite{LoustoS1992} yield to the conclusion that the resulting collisional and 
particle interacting effect can be  well described by the field felt by one particle in the effective gravitational curved background produced by all the others. Here, we consider the (non perturbative) curved space-time background from the beginning. This can be thus viewed as the effective background field felt by one particle produced by the particle interactions of all the others. 

In the early Planckian and trans-Planckian phases, namely the nearest possible ones to the universe origin, it is natural to consider the $\Lambda$ or vacuum dominance background as we consider which is also motivated by a description of the origin of the  universe "from nothing".   Indeed, the  $\Lambda$ 
 or vacuum  dominance background in the early phases could be considered as formed as a 
 condensate from such particle interactions. In  
 summary, the effective result of such particle 
 interactions is to produce the curved background.

\section{The Harmonic Oscillator and the Cosmological Constant}

For simplicity and physical insight we consider the case of just one 
particle, Eqs.(6.5) and (6.6) yield:
\begin{equation} \label{motion}
{\ddot{\vec X}}=\frac{\Lambda}{3} \, {\vec X} 
\end{equation}
This is an {\it harmonic oscillator} equation with imaginary frequency and {\it oscillator constant} $\kappa_{oscill}$:
\begin{equation} \label{constantka}
{\ddot{\vec X}} = -\kappa_{oscill} {\vec X}, \qquad
\kappa_{oscill} = \omega^2 m , \qquad \omega = \sqrt{\frac{\Lambda}{3m}},
\end{equation}
with the solution,
\begin{equation} \label{tray}
{\vec X}(t) = {\vec X}(0)\; \cosh Ht + \frac1{H} \; {\dot{\vec
X}}(0)\; \sinh Ht \; ,
\end{equation}
where 
$$ H \equiv \sqrt{ \Lambda / 3 }$$
The particle runs away exponentially fast in time. The Hubble constant $H^2$ is the constant of the oscillator
\begin{equation} \label{constantkb}
\kappa_{osc} = H^2,\quad H = \omega \sqrt{m} 
\end{equation}
the oscillator length $l_{osc}$ being 
$$ l_{osc} = \sqrt{3/\Lambda}, \qquad H = c/l_{osc} = \kappa \equiv \text{surface~gravity}$$

The length of the oscillator is the Hubble radius and the Hubble constant is the surface gravity 
of the universe (similar to the black hole surface gravity, the inverse of the black hole 
radius).

\medskip

The non-relativistic or weak field newtonian 
results reproduce very well the full space-time 
relativistic effects in the presence of the 
cosmological constant. 
The exact solution of the Einstein equations for 
the energy-momentum tensor eq.(\ref{tmunu}) with $ 
\rho = 0 $ is the de Sitter universe.
It must be stressed that the non-relativistic 
trajectories Eq.(\ref{tray}) exhibit the same 
exponential runaway behaviour of the exact 
relativistic 
geodesics in de Sitter space-time.
The non-relativistic approximation  keeps the 
essential features of the particle motion in de 
Sitter space-time \cite{Tolman}, 
\cite{dVSiebert}, 
\cite{dVSanchez}. 

\medskip

The description of de Sitter space-time as an 
(inverted) harmonic oscillator appears either in a relativistic or in a non-relativistic 
consideration. This stems from its geometrical 
hyperbolic 
description: $- T^2 + X^2 + X_i^2 + Z^2 = L^2$ 
as an hyperboloid embedded in a flat Minkowski 
space-time with one more spatial dimension. The 
propagation equations of particles, waves, 
fields and strings in de Sitter space-time all 
reflect the de Sitter space-time connection to 
the inverted harmonic oscillator, namely in all 
these cases, a term of the form and 
sign of the inverted oscillator 
Eq.(\ref{hamil}), Eq.(\ref{motion})          
does appear, e.g see for example 
\cite{dVSanchez}, 
\cite{GuthPi},\cite{Albrecht},
\cite{SanchezNPB1987}. In particular, in several regimes, e.g. asymptotically for $t \rightarrow \pm \infty$ or for $\Lambda$ dominance with respect to other field parameters (as masses and couplings), the propagation equations in de Sitter space-time (cosmic time for instance) reduce to
$$
\ddot{\chi} - \nu^2 H^2 \chi= 0, \;\;\quad  \nu^2 \equiv \nu^2 (m^2, H^2, \xi).  
$$
In general,  $\nu^2 \equiv \nu^2 (m^2, H^2, \xi)$, 
and its sign depend on the relationship between  $H$, $m$ and the couplings $\xi$. For instance, for inflation, $H^2$ dominates over $m^2$ and the couplings, and $\nu^2$ is positive. Squeezed states are characteristic of this propagation, e.g. for quantum fields see for example references \cite{GuthPi},\cite{Albrecht}.

\medskip

We summarize in the following our main results 
allowing to describe de Sitter (and Anti de 
Sitter) space-time as a classical and quantum 
harmonic oscillator:

\begin{itemize}
\item{The motion of a particle in an harmonic 
oscillator potential corresponds to the particle 
motion in the non-relativistic limit of 
a constant curvature space-time. The harmonic 
oscillator with an imaginary frequency, namely the 
inverted oscillator for $\Lambda > 0$ 
corresponds to  de Sitter space-time;  the real 
frequency normal oscillator $\Lambda < 0$ describes anti-de Sitter space-time, and the 
free motion is flat Minkowski space-time $\Lambda = 0$.}

\item{The constant of the oscillator is the 
cosmological constant, as shown by 
Eq.(\ref{constantka}), which is  the Hubble 
constant $ H^2$  or surface gravity squared  Eq.(\ref{constantkb}).}

\item{For the {\it classical} harmonic oscillator, 
the phase space is the classical one, and the 
algebra of the $(X, P)$ variables or $(X, T )$ 
variables is commuting. 
The classical Hamiltonian is $2 H_{osc} = X^2 + P^2$  or $2 H_{inv-osc} = X^2 - P^2$ for the inverted oscillator, in light-cone variables $2 UV= 2VU $.
The light-cone structure $X^2 - T^2 $ is the 
classical known one, there is no difference with 
the Minkowski light-cone structure of special relativity. 
Upon the identification $P = T$  the classical 
commuting $(X,T)$ variables of Minkowski space-time and its invariant distance $s^2 = X^2 -T^2$ correspond
to a classical phase space $(X, P)$ and Hamiltonian $s^2 = 2H_{inv-osc} = X^2 - P^2$ which is the (inverted) harmonic oscillator Hamiltonian.}

\item{The non relativistic approximation describes very well the essential properties 
of the constant curvature -de Sitter or anti de Sitter- geometries and captures its physics.
Thus, the classical non-relativistic de Sitter invariant space-time, or the anti-de Sitter space-time, and the Minkowski Poincare-invariant space-time 
all three describe  special relativity. We see that this reaches from another approach and motivation, the fact that a constant curvature space-time 
describes special relativity, as refs \cite{Bacry} \cite {Dyson}, or the so called "triply relativity" $\Lambda > 0$, $\Lambda < 0$ and $\Lambda = 0$.} 

\item{For the {\it quantum} harmonic oscillator, the quantum zero point energy bends 
the light hyperbolic cone generators $X^2 -T^2 = 1$
and therefore the space-time is curved: de Sitter (or anti de Sitter)
space-time. And, as it is known, the non-relativistic and relativistic de Sitter space-times are very similar.}

 \item {Upon the identification $T = P$, the classical non commuting coordinates $(X, T)$ of Minkowski space-time 
and its distance $ s^2 = X^2 -T^2$ are the classical non commuting phase space $(X, P)$  
and classical quadratic form  $2H_{inv-osc}= X^2 - P^2 \rightarrow s^2$, which  
is the harmonic oscillator Hamiltonian. And this is too the Hamiltonian of a particle in a constant curvature (cosmological constant) de Sitter or anti de Sitter space (in its non-relativistic limit).}

\item{ Explicitely: The $(a, a^+)$ creation and annihilation operators are the {\it light-cone} type quantum coordinates of the phase space $(X, P)$: 
$a = \;(X + i P)/\;\sqrt{2},
\quad a^+ = \;(X - i P)/\;\sqrt{2}$.
The temporal variable $T$ in the space-time configuration $(X, T)$ is like the momentum in phase space $(X, P)$. The identification $P = T$ yields: 
\be 
X = \; (a + a^+) /\;\sqrt{2} , \qquad  
T = \;(a  - a^+ )/i\sqrt{2} \;,  \qquad [a, a^+] = 1
\ee
$$
2 X^2 = [(2 a^+ a + 1) + (a^2 + a^{+ 2})], \qquad 
2 T^2 = [(2 a^+ a + 1) - (a^2 + a^{+ 2})]
$$
with the algebra:
$$
H_{osc} = (X^2 + T^2) = (2 a^+ a + 1), \; \qquad 
H_{inv-osc} = (X^2 - T^2) = (a^2 + a^{+ 2}),
$$
\be
[X, T] = i, \qquad \; [H_{inv-osc}, X]  = 2iT, \qquad \;  [H_{inv-osc}, T]  = 2iX,   
\ee
$ a^+\;a = N $ being the number operator.}

\item{In other words: The non-relativistic cosmological constant (de Sitter or Anti de Sitter) space-time, 
the harmonic oscillator phase space and Minkowski space - time are in correspondence one into another. The line element in Minkowski space-time in 
$D$ space-time dimensions 
$s^2 = X^2 - T^2 + X_j^2$ is equal to the  (non relativistic) harmonic oscillator Hamiltonian $2H_{inv-osc} = X^2 - P^2 + X_j^2$.
Thus, there are the three possibilities for special relativity.
The interesting point in our studies is that the {\it quantum} harmonic oscillator algebra describes the {\it quantum non-commuting space-time} 
structure.} 

\item{Upon the identification $T = P$, the de Sitter hyperboloid Eq.(\ref{dS}) yields :
\begin{equation}
X^2 - P^2 + X_j^2 + Z^2 = L_{QH}^2, \qquad j= 2, 3, ...(D-2)
\end{equation}
corresponding to a (inverted) harmonic oscillator $(X,P)$ embedded 
in a Minkowski space of $(D-2 + 2) = D$ spatial dimensions, ie a  Minkowski space-time of $(D+1)$ space-time dimensions.} 
\end{itemize}

\section {Quantum Discrete Levels of the Universe}

Let us  go  beyond  the  classical-quantum  duality  of  the  space-time  recently  discussed  and promote the space-time 
coordinates to quantum non-commuting operators.
As we have seen, comparison to the harmonic oscillator $(X, P)$ variables and global phase space
is enlighting: The hyperbolic phase space here $(X, P = T)$ describes the  
hyperbolic quantum space-time structure and generates the quantum light cone.
The classical Minkowski space-time null generators $X = \pm T$ {\it dissapear} 
at the quantum level due to the relevant 
$[X,T]$ conmutator which is {\it always} non-zero. A  new quantum Planck scale vacuum region emerges.
In the case of the Rindler and Schwarzshild-Kruskal space-time structures, 
the four Kruskal regions merge inside a   single quantum Planck scale region 
\cite{Sanchez2019}, \cite{Sanchez2}. 

\medskip

The quantum space-time structure consists of {discrete levels} 
of odd numbers 
\be
X^2_n = (2n + 1)\;\;\;,   T^2_n = (2n + 1) \quad
\mbox{(in Planck units )},\;\; n = 0, 1, 2.... 
\ee
$(X_n, T_n)$ and the {\it mass levels} being $\sqrt {(2n + 1)}$,\; $n = 0, 1, 2....$

\medskip

The Planck scale level $(X , T)(n=0)  =  1 $  is the fundamental $(n=0)$ level from which  the space-time levels $(X_n , T_n)$ go to the quantum (low $n$) levels and to the semiclassical  and classical (large $n$) levels. Asymptotically, for very large $n$ the space-time becomes continum. 

\medskip

In terms of variables $(x_{n\pm}, t_{n\pm})$, covering only one: the pre-Planckian or the post-Planckian phase, the space-time discrete levels read:
\be
x_{n\pm} = [\;\sqrt{ 2n +1 } \pm \sqrt{2n}\;]  
\ee 
\be
t_{n\pm} =  [\sqrt{2n+1} \pm \sqrt{(2n+1) + 1/2}\;],
\ee
$$x_{n = 0}\;(+) = x_{n = 0}\;(-) = 1: \mbox{Planck scale}$$ 
The low $n$, intermediate, and large $n$ levels describe respectively the quantum, semiclassical and classical behaviours, interestingly enough 
the $(\pm)$ branches consistently reflect the classical-quantum duality properties.

$(X_n,T_n) , (x_n, t_n) $ are given in Planck (length and time) units. In terms of the global 
quantum gravity dimensionless length ${\cal L} = L_{QH}/l_P$ and mass ${\cal M} = M_{QH}/m_P$, 
Eqs. (\ref{dimless}) or the local ones $x = m/m_p$,  translate into the discrete mass levels:
\be \label{Ln} 
{\cal L}_n = \sqrt{(2n + 1)}= {\cal M}_n \;\;\; n = 0, 1, 2,....
\ee
\be
L_{QHn \;\;n >>1}\;= 
\; l_P \; [\;\sqrt{2\;n}\; + \frac{1}{2\sqrt{2\;n}} + O(1/n^{3/2})\;]  
\ee
\be
M_{QHn \;\;n >>1}\;= 
\; m_P \; [\;\sqrt{2\;n}\; + \frac{1}{2\sqrt{2\;n}} + O(1/n^{3/2})\;]  
\ee
 The above Eqs for $L_{QHn}$, $M_{QHn}$ yield the levels for $L_{Hn \pm}$ and $M_{Hn \pm}$:
\be
L_{Hn \pm} =  [\; L_{QHn} \pm \sqrt{ L_{QHn}^{2} - l_P^2 }\;] 
\ee
\be
M_{Hn \pm} =  [\; M_{QHn} \pm \sqrt{ M_{QHn}^{2} - m_P^2 }\;] 
\ee
The condition $L_{QHn} \geq l_P$, $M_{QHn} \geq m_P$ consistently corresponds to the whole spectrum $n\geq 0$,
the lowest level $n=0$ being the Planck mass and length:
\be
L_{Hn \pm} = l_P \; [\;\sqrt{ 2n +1 } \pm \sqrt{2n}\;] \quad  \; \mbox {for all $n = 0, 1, 2,...$}
\ee
\be
M_{Hn \pm} = m_P \; [\;\sqrt{ 2n +1 } \pm \sqrt{2n}\;] \quad  \; \mbox {for all $n = 0, 1, 2,...$}
\ee

The mass and radius of the universe $M_H$, $L_{H}$ have discrete levels $L_{Hn \pm}, M_{Hn \pm}$, from the most fundamental one $(n = 0)$, 
going to the semiclassical (intermediate $n$), to the classical ones (large $n$) which yield a continumm classical universe as it must be. 
This is clearly seen from the mass level $M_{Hn \pm}$ expressions (and similarly for the radius levels). Explicitely:
\be
\quad M_{H (n=0)+} = M_{H (n=0)-} = M_{QH (n=0)} = m_P,  \quad  \; n = 0 : \mbox {planck mass}  
\ee
\be M_{Hn +} = m_P \; [\;2\sqrt{2\;n}\; - \;\frac{1}{2\sqrt{2 n}}\; 
+ \;O (1/n^{3/2})\;] , \;\;\;\quad  \mbox{large $n$ : branch $(+)$ : masses $> m_P$}
\ee  
\be M_{Hn-} = \frac{m_P}{ 2 \sqrt{2\; n}}\; + \;O(1/n^{3/2}) , \quad \;\;\;
\mbox{large $n$ : branch $(-)$ : masses $< m_P$ } 
 \ee
Large $n$ levels are semiclassical tending towards a classical continuum space-time. Low $n$ are quantum, the lowest mode ($n = 0$) being the Planck scale. 
Two dual $(\pm)$ branches are present in the local variables ($\sqrt{2n+1}\pm \sqrt{2n}$) reflecting the duality of the 
large and small $n$ behaviours and covering the {\it whole} spectrum:
from the largest cosmological masses and scales in branch $(+)$
to the quantum smallest masses and scales in branch $(-)$ passing by the Planck mass and length. 

\section {Quantum Discrete Levels of the Hubble Constant}

Eqs.(\ref{Ln}) yields the (dimensionless) quantum levels for the total: Hubble constant, vacuum energy and constant curvature:
\be \label{Hn}
{\cal H}_n = \frac{1}{\sqrt{(2n + 1)}}, \;\;  \qquad  
\mathbf{\Lambda}_n = \frac{ 1 }{(2n + 1)}, \;\; \qquad {\cal R}_n = \frac{ 1 }{(2n + 1)}
\quad n = 0, 1, 2, ...
\ee
\be
n = 0: \;\; {\cal H}_0 = 1, \qquad \mathbf{\Lambda}_0 = 1, \qquad {\cal R}_0 = 1: \quad \mbox {Planck scale (dimensionless)}  
\ee
\be
H_{QH \;(n=0)} = \frac{c}{l_P} = h_P, \; \quad  \Lambda_{QH \; (n=0)} = \lambda_{P} \; \quad
R_{QH \; (n=0)} = 4 \lambda_{P}:
\;\; \; \mbox {Planck scale values}  
\ee
And for the gravitational entropy:
$$S_n = (2n + 1)\; \;\; \mbox{in Planck units \; $s_P= 4\pi$} 
$$
The lowest $n = 0$ level corresponds to the fundamental Planck scale values $(h_P, \lambda_P, 4\lambda_P, s_P)$ for the Hubble constant, 
cosmological constant, constant curvature and gravitational entropy respectively. Let us analyze now the implications of these results and 
the general picture which they arise. 

\medskip

In {\bf the post-Planckian universe $t_P \leq t \leq t_{today} = 10^{61} t_P$:}
We see that the physical magnitudes as the Hubble radius, vacuum energy density, constant curvature, entropy start at the Planck scale: 
the zero level $(n = 0)$. As $n$ increases, the universe radius, mass and entropy increase, the Hubble constant, curvature and vacuum energy 
{\it consistently decrease} and the universe {\it classicalizes}. The decreasing  with $n$ of these quantities is given by Eq.(\ref{Hn}), 
and for large $n$, $H_{n}$, $\Lambda_{n}$ and ${\cal R}_{n}$ {\it classicalize} as:
\be
{\cal H}_{n >> 1} = \frac{c}{l_P \sqrt{2n}}\; [\;1 - O \; (\;\frac{1}{2n}\;)\;] \; << 1
\ee
\be
\mathbf{\Lambda}_{n >> 1} = \frac{3\; c^2}{l_P^2\; (2n)}\; [\; 1 -  O \; (\;\frac{1}{2n}\;)\; ] \; << 1
\ee
\be
{\cal R}_{n >> 1} = \frac{12\; c^2}{l_P^2\; (2n)}\; [\; 1 - O \; (\;\frac{1}{2n}\;)\; ] \; << 1,
\ee
\medskip
{\it precisely accounting for the low classical values of $H$ and $\Lambda$ in the universe today which is a classical, large and dilute universe}. 
The present universe values $H_{today}= 10^{-61}$, 
$\rho_{\Lambda} = 10^{-122}$ correspond to a large  $n$-level $n = 10^{122} \equiv n_{today}$.

\medskip

More generally, in the {\bf post-Planckian universe}: 
$t_P \leq t \leq t_{today} = 10^{61} t_P$, 
Eq.(\ref{Hn}) yields the quantum $n$-levels:
\be
n = \frac{1}{2}\;(H_{n} ^{-2} - 1):  \;\;\;\; t_{(n=0)}= t_P \leq t_n \leq  t_{n \;today} = 10^{61} t_P
\ee
Thus, the more characteristic evolution values from the Planck time $t_P$ till today: 
\be
h_P,..., H_{inf},..., H_{cmb},..., H_{reion}, ...,H_{today} \;,
\ee
corresponds to the $n$-levels: 
\be
n = 0, 1, 2,... n_{inf} = 10^{12},...n_{cmb} = 10^{114}, ... n_{reoin} = 10^{118},...n_{today} = 10^{122}
\ee
and the {\it discrete} $ H_n$, $\Lambda_n$ and $S_n$ values:  
\be
H_n = 1,\; 0.577,...H_{n,inf} = 10^{-6},...
H_{n,cmb} = 10^{-57},... H_{n,reoin} = 10^{-59},... H_{n,today} = 10^{-61} 
\ee
\be
\Lambda_n = 1, \; 0.333,... \Lambda_{n,inf} = 10^{-12},... 
\Lambda_{n,cmb} = 10^{-114},...\Lambda_{n,reoin} = 10^{-118},...\Lambda_{n,today} = 10^{-122} 
\ee
\be
S_n = 1, \; 3,... S_{n,inf} = 10^{12},... S_{n,cmb} = 10^{114},...S_{n,reoin} = 10^{118},...S_{n,today} = 10^{122} 
\ee

In {\bf the pre-Planckian or precursor phase}, namely, the trans-Planckian phase:
\be
 10^{-61}t_P \;\leq \;t_n \;\leq \; t_P \; (n = 0),
 \ee
the quantum $n$-levels for $ H_{Qn}$, $\Lambda_{Qn}$, $S_{Qn}$ Eqs (\ref{dualQH}) are:
\be
H_{Qn} = \sqrt{2n + 1} \;, \; \; \;\Lambda_{Qn} = (2n + 1) \;, \;\;\; S_{Qn}= \frac {1}{(2n + 1)}, \;\; n = 0, 1, 2, ....
\ee
Thus:
\be
n = \frac{1}{2} \;(H_{Qn} ^{2} - 1),  \;\;\;\; 10^{-61}t_P \leq t_n \leq t_P \; (n=0)
\ee
and the more characteristic values in this phase, namely:  
\be h_P,...H_{Qinf},...H_{Qcmb},...H_{Qreion}, ... H_{Qtoday} \equiv H_{far\;past}, \ee
correspond to the $n$ - level values:
\be
n = 0, 1,...n_{Qinf} = 10^{12},...n_{Qcmb} = 10^{114}, ...n_{Qreoin} = 10^{118},...n_{Qtoday} \equiv n_{far\;past}= 10^{122} 
\ee
And the $H_{Qn}$, $\Lambda_{Qn}$ and $S_{Qn}$ levels have the values:  
\be
H_{Qn} = 1,\; 1.732,...H_{Qinf} = 10^{6},...
H_{Qcmb} = 10^{57},...H_{Qreoin} = 10^{59},... H_{Qtoday} = 10^{61} 
\ee
\be
\Lambda_{Qn} = 1, \; 3, ... \Lambda_{Qinf} = 10^{12},... 
\Lambda_{Qcmb} = 10^{114},...\Lambda_{Qreoin} = 10^{118},...\Lambda_{Qtoday} = 10^{122} 
\ee
\be
S_{Qn} = 1, \; 0.333,...S_{Qinf} = 10^{-12},... 
S_{Qcmb} = 10^{-114},...S_{Qreoin} = 10^{-118},...S_{Qtoday} = 10^{-122} 
\ee

The whole picture is described at the end of Section X including {\bf both the pre-Planckian and post-Planckian phases}, and the complete discrete spectrum of levels from the far past to today level.
The universe pre-Planckian phase, namely the quantum precursor phase is the setting of the physically meaningful quantum trans-Planckian energies.
In the post-Planckian (semiclassical and classical)  eras, {\it no} trans-Planckian energies are present: only mathematically or artificially (non physical) trans-Planckian energies could be generated in the present universe. This is a direct consequence of the classical-quantum gravity dual relations Eqs (2.2), (2.3), (2.4),  which apply to any physical relevant magnitude, and in the  respective domains as discussed in  Sec II  and refs [1], [2], [3].  The transplanckian energy domain remains in the phase of the universe totally before the Plank time $t_P$, eg in a totally quantum gravity domain, while the energies in the universe after the Planck time turn out smaller than the Planck energy (semiclassical or semiquantum gravity and classical gravity) as determined by the  classical-quantum gravity duality (CQGD) relations. However, signals or observables from the quantum precursor phase are present in the classical and semiclassical universe, the most known being  inflation and the present dark (vacuum) energy.
   
\medskip

Consistently, the pre-Planckian phase covering $10^{-61} t_P \leq t \leq t_P$,  provides too the two dual: $(+)$ and $(-)$ branches, as it must be:
\be
H_{n \pm} = h_P \; [\;\sqrt{2n + 1} \pm \sqrt{2n}\;] \qquad n = 0, 1, 2....
\ee
\be
H_{n=0} = h_P :\quad \quad \mbox{Planck scale value} 
\ee
\be
H_{n +,\;\; n>>1} = h_P \; [\;2\sqrt{2\;n}\; - \;\frac{1}{2\sqrt{2 n}}\; 
+ \;O\;(1/n^{3/2})\;]\; >> 1 \;\;\;\quad  \mbox{large $n$: branch (+)}
\ee  
\be H_{n -,\;\; n>>1} = \frac{h_P}{2 \sqrt{2\; n}}\; + \; O\; (1/n^{3/2})\; << 1 \;\;\;\quad
\mbox{large $n$ : branch (-) } 
\ee
And for the universe radius levels $L_{Hn}$:
\be
\quad L_{H (n=0)+} = L_{H (n=0)-} = L_{QH (n=0)} = l_P  \quad  \; n = 0 : \mbox {planck length}  
\ee
\be L_{Hn +,\;\; n>>1} = l_P \; [\;2\sqrt{2\;n}\; - \;\frac{1}{2\sqrt{2 n}}\; 
+ \;O (1/n^{3/2})\;]\; >> 1 \;\;\;\quad  \mbox{ large $n$ : branch (+) }
\ee  
\be L_{Hn-,\;\; n>>1} = \frac{l_P}{ 2 \sqrt{2\; n}}\; + \;O(1/n^{3/2})\; << 1 \quad \;\;\;
\mbox{large $n$ : branch (-)} 
 \ee

\medskip

The same expressions hold for the mass levels $M_{Hn} (\pm)$; the vacuum levels $\Lambda_n (\pm)$ and the gravitational entropy $S_n (\pm)$ levels follow from them. 

\medskip

The quantum levels cover {\it all} the range of scales from the largest cosmological scales and time $10^{61} t_P$ today to the smallest one $10^{-61} l_P$ 
in the extreme past $10^{-61} t_P$ of the precursor or trans-Planckian phase, passing through the Planck scale $(l_P, t_P)$, covering the two phases: 
post and pre Planckian phases respectively.
The quantum mass levels are associated to the quantum space-time structure. 
Quantum mass levels here cover {\it all} masses $10^{-61} m_P \leq  M_n \leq 10^{61} m_P$ of the universe phases. The two {\it dual mass} 
branches ($\pm$) correspond to the larger and smaller masses with respect to the Planck mass $m_P$ respectively, they cover the 
{\it whole mass range} from the Planck mass in branch $(+)$ until the 
largest cosmological masses, and from the smallest masses in branch $(-)$, the pre-Planckian phase, til near the Planck mass.
As $n$ increases, masses in the branch $(+)$ {\it increase} (as $2 \sqrt{2n}$). 
Masses in the branch $(-)$, the very quantum one, {\it decrease} in the large $n$ behaviour, precisely as $ 1 /(2 \sqrt{2n})$, 
large $n$ are very excited levels in this branch, {\it consistently} with the fact that this branch is the dual of branch $(+)$. 

\section {The Snyder-Yang Algebra and Quantum de Sitter Space-Time}

The space-time coordinates in the Planckian and super-Planckian domain are no longer commuting, but they obey non-zero commutation relations: 
The concept of space-time is replaced by a quantum algebra. The classical space-time is recovered from the quantum algebra as a particular 
case in which the quantum space and time coordinate operators become the classical space-time continumm coordinates (c-numbers) with all 
commutators vanishing and the discrete spectrum becomes the classical continumm space-time. 

\medskip

 Here the quantum space-time description is reached {\it directly} from the quantum non-commuting space- time coordinates and not through the 
 quantization procedure of the classical gravitational field. This is so because the gravity field is itself a classical concept which loose meaning 
 at the Planck scale. The space-time (the arena of events) is a classical concept which is more direct to extend to, or to replace by, 
 {\it a quantum algebra} of space-time position and momenta
$$ 
[X_i, X_j ] = i M_{ij} 
$$

The Snyder algebra is a Lorentz covariant deformation of the Heisenberg algebra, 
where the position operators are non-commuting and have discrete spectra \cite{Snyder} 
soon extended by Yang \cite{Yang} to include one more length parameter. It describes 
a non-commutative discrete space-time compatible with Lorentz-Poincare symmetry. 
The discrete position spectra, representations of the algebra imply a discrete
space description of space.

\begin{itemize}
\item{The Snyder algebra is precisely a description of a 4D constant curvature
 space of momenta, this corresponds to a de Sitter hyperboloid embedded in a 5D Minkowski momentum space.
 In the space of 5D momenta $p_A$, this includes precisely 
 the motion of a particle of mass $m$ and momentum on the de Sitter momentum hyperboloid $\eta^{AB} p_A p_B = m^2$.}

\item{In geometric terms, the Snyder quantized space-time is a projective 
geometry approach to the phase space or momentum de Sitter space in which the space-time 
coordinates are identified with the 4-translation generators  of the $SO (1,4)$ de Sitter 
group (and are therefore non-commutative), and with other operators as the angular momentum in $SO
(1,3)$.}

\item{In projective or Beltrami coordinates, the  Euclid, Riemann and Lobachevsky 
spaces \cite{Guo} corresponding to zero, positive and negative spatial curvature respectively, are upon Wick 
rotation the Minkowski, de Sitter and Anti de Sitter space-times with the invariance groups 
 $ISO (1,3), SO(1,4), SO(2,3)$ respectively.}
\end{itemize}
In D dimensions, the Lorentz-covariant Snyder-Yang quantum algebra follows from the Inonu-Wigner 
 \cite{Wigner} group contraction of the $SO(D-1,1)$ algebra with the generators:  
\be \label{sigma}
\Sigma_{AB} = i (q_A \partial_{q_B} - q_B \partial_{q_A} ),
\ee
$\Sigma_{AB}$ live on the $(D+2)$ parameter space $q_A$ (hyperboloid) which satisfies
\be \label {ds}
- q_0^2 + q_1^2 + ... + q_{D-1}^2 + q_a^2 + q_b^2 = L^2
\ee
\be
A = (\mu,  a, b); \;(\mu = 1, 2, ...D);\;(a,b) \mbox{being extra space dimensions, and $q_0 \equiv q_D$}.
\ee
The $D$-dimensional operators $(X_\mu, P_\mu, M_{\mu\nu})$: space-time operator $X_\mu$, momentum operator $P_\mu$, angular momentum operators 
$M_{\mu\nu}$ and the completing operator $N_{ab}$ are all defined by the generators 
$\Sigma_{\mu a}$ Eq.(\ref{sigma}) as following:
\be \label {oper}
X_\mu \equiv l_P\Sigma_{\mu a}, \qquad P_\mu \equiv (\hbar / L) \Sigma{\mu b}, \qquad
M_{\mu\nu} \equiv \hbar \Sigma_{\mu \nu}, \qquad N_{ab} \equiv (l_P / L) 
\Sigma_{a b}.
\ee
This set of operators ($X_\mu, P_\mu, M_{\mu\nu}, N $) satisfy the contracted algebra of $SO(D+1,1)$, namely the quantum Yang-Snyder 
space-time algebra:
\be \label {XPM}
[ X_\mu, X_\nu ] = -i (l_P^2/ \hbar) M_{\mu\nu}, \qquad [ P_\mu, P_\nu ] = -i (\hbar / L^2) M_{\mu\nu},
\ee
\be \label {XPN}
[ X_\mu, P_\nu ] = -i \hbar N \delta_{\mu\nu}, \qquad [ X_\mu, N ] = i (l_P^2 /\hbar) P_{\mu}, 
\qquad [ P_\mu, N ] = - i (\hbar /L^2) X_{\mu}
\ee
And the operators $ M_{\mu\nu}$ satisfy angular momentum's type  relations:
\be \label {MM}
[ M_\mu, M_\nu ] = - i (l_P^2/ \hbar) M_{\mu\nu}
\ee
{\bf Classical-quantum duality in the Snyder-Yang algebra}: The Snyder-Yang algebra contains 
two parameters $(a, L)$: small scale parameter $a$ and large scale parameter $L$ which in our context are 
naturally the Planck length $l_P$ and the universe radius $L_H$.
Our complete (classical and quantum) radius $L_{QH}$ Eq.(\ref{LQH})  {\it contains intrinsically the both lengths}, the classical length $L_H$ 
and its quantum dual 
$L_Q$ (Compton radius of the universe), and provides a basis for a framework naturally free of infrared and ultraviolet divergences:  
\be
a \equiv l_P, \;  L \equiv L_{QH} = L_H + L_Q = l_P \;(\frac{L_H}{l_P}  + \frac{l_P}{L_H})
\ee \label {aR}
We see that the Snyder-Yang algebra with the complete length $L_{QH} (l_P, L_H)$ as a parameter provides a quantum operator realization 
of the complete
(classical and quantum) de Sitter universe, including the quantum early and classical late de Sitter phases duals of each other. 
This provides further description of  the pre-Planckian and post-Planckian de Sitter phases, within a group-theory realization of 
the quantum discrete de Sitter space-time and of classical-quantum gravity duality. 

Finally, let us mention as an example of the different classical and quantum de Sitter phases: the cosmological vacuum energy, the most direct candidate to the dark energy today
, \cite{Riess},\cite{Perlmutter},\cite{Schmidt},\cite{WMAP1},\cite
{WMAP2},\cite{DES},\cite{Planck6}, for which the observed value is:
\begin{equation} \label{rhovalue}
\rho_\Lambda = \Omega_\Lambda \rho_c =  3. 28 \; 10^{-11} (eV)^4 = (2.39 \; meV)^4, \qquad  m eV= 10^{-3} eV
\end{equation}
corresponding to $ h = 0.73, \quad \Omega_\Lambda = 0.76 , \quad H = 1.558 \; 10^{-33} eV$.  The CMB data yield the values \cite{Planck6}: 
\begin{equation} \label{Hvalue}
H = 67.4 \pm 0.5 \; Km \; sec^{-1}\; Mpc^{-1}, \quad \Omega_\Lambda h^2 = 0.0224 \pm 10^{-4}
\end{equation} and \begin{equation} \label{Omegavalue}
\Omega_\Lambda = 0.6847 \pm 0.0073, \quad  \Omega_\Lambda h^2 = 0.3107 \pm 0.0082,
\end{equation} 
which implies for the cosmological vacuum {\bf today}:
\begin{equation} \label{Lambdavalue} 
\Lambda = (4.24 \pm 0.11) \; 10^{-66}\; (eV)^2 = (2.846 \pm 0.076) \; 10^{-122}\; m_P^2
\end{equation} 
The density $\rho_\Lambda$ associated to $\Lambda$ Eq.(\ref{rhovalue}) 
is precisely:
\begin{equation} \label{rhoLambda} 
\rho_\Lambda = \Lambda /8 \pi G = \rho_P \left(\Lambda/\lambda_P\right),
\end{equation} 
where the planck scale values $\rho_P,\lambda_P$ are: 
$\rho_P = \lambda_P/8 \pi G, \quad \; \lambda_P = \ 3 h_P^2$. 
The quantum vacuum value expected from microscopic particle 
physics is evaluated to be $\Lambda_Q \approx 10^{122}$.

{\bf Crossing the Planck scale}. The two values: $(\Lambda,\Lambda_Q)$, 
refer to the same concept of vacuum energy but they are in two huge different vacuum states and two huge different cosmological epochs: classical state 
and classical dilute epoch today for $\Lambda$ observed today with the most classical levels, and quantum state 
and quantum very early epoch with the most excited levels for the quantum mechanical 
transplanckian value $\Lambda_Q$. The classical value today $\Lambda = 3H^2$ corresponds to the classical universe today of classical rate $H$  
and classical cosmological radius $L_H = c/ H$. The quantum mechanical 
value $\Lambda_Q = 3 H_Q^2$ corresponds to the early quantum universe of quantum rate $H_Q$ and
quantum radius $L_Q = l_P^2/L_H = \hbar/M_H c$ which is  exactly the quantum dual of the classical
horizon radius $L_H$: $L_Q$ is {\it precisely} the quantum Compton length of the universe for the
gravitational mass $M_H = L_H c^2/G$.

{\bf Two extremely different physical conditions and gravity regimes}. This is a realistic, clear and precise illustration of the 
{\it physical classical-quantum 
duality} between the two extreme Universe scales and gravity regimes or phases {\it through the Planck scale}: the dilute state and horizon size 
of the universe today on the one largest known side, and the trans-Planckian scales and highest density state on the smallest side: size, mass, 
and their associated time (Hubble rate) and vacuum energy density 
($\Lambda, \rho_\Lambda$) of the universe {\it today} are truly
{\it classical},  while its extreme past at $10^{-61}\; t_P$ = $ 10^{-105}$ sec deep inside the trans-Planckian domain of extremely 
small size and high vacuum density 
value ($\Lambda_Q, \rho_Q$) are truly {\it quantum and trans-Planckian}.  
This manifests the {\it classical-quantum or wave-particle duality} between  the classical macroscopic 
(cosmological) gravity physical phase and 
the quantum microscopic particle physics and trans-Planckian phase
through the {\it crossing} of the Planck scale, Planck scale duality in short.

{\bf An unifying picture:}
Starting from the earliest past quantum era from $10^{-61}\; t_P$ to $t_P$, with  the quantum 
excited level $n = 10^{122}$, the entropy $S_{Qn}$ increases in discrete levels $ s_P/(2n+1)$ from its 
extreme small value $S_{Q} = 10^{-122} \;s_P$ at the earliest time $ 10^{-61}\; t_P$ 
till for instance its quantum inflation value $10^{-12} \;s_P$, $( n_{Qinfl} = 10^{12})$, at time $10^{-6}\; t_P$, to its Planck 
value $(n = 0): S_Q = s_P = \pi \kappa_B$ at the Planck time $t_P$, the {\it crossing scale}, after which it 
goes to its semi-classical and classical levels $(2 n + 1) s_P$, e.g. inflationary  value $S_{\Lambda \;inflation} = 10^{12} \;s_P$ , $(n = 10^{12})$ at 
the classical inflationary stage at $10^{6} \; t_P$ 
and it follows {\it increasing and classicalizes} till the most 
classical level today $n = 10^{122}$: $S_{\Lambda} = 10^{122} \;s_P$ at the present time $10^{61}\; t_P$.  And as far as the universe will 
continue expanding its horizon as $l_P \sqrt{(2n + 1)}$, $S_{\Lambda n}$ will continue increasing as $(2n + 1)$.

The {\it total} $Q\Lambda$ gravitational entropy (for the whole history) is the sum of the three values 
above discussed 
corresponding to the three regimes: classical $\Lambda$, quantum $Q$ and Planck 
values (subcript $P$).
In the past remote and more quantum (Q) eras: $10^{-61} \;t_P \leq  t \leq t_P$, 
the Planck entropy value $(n = 0)$: $s_P = \pi \kappa_B$ dominates $S_Q$. In the classical eras:
$t_P \leq  t \leq 10^{61} t_P$, the today entropy value $(n = 10^{+122}): S_\Lambda = 10^{+122} s_P$ dominates.

{\bf The whole picture is depicted in Figure (1)}, where: $\Lambda$ refers to the cosmological constant (or associated Hubble-Lemaitre constant H) in the 
classical gravity phase. Q means quantum, P means Planck scale, Planck's units, natural to the 
system, greatly simplify the history. (The complete history is a theory of pure numbers). Each stage is characterized by the set of main physical gravitational quantities: ($\Lambda$, density $\rho_\Lambda$, 
size $L_\Lambda$, and gravitational entropy $S_\Lambda$). In the quantum trans-Planckian phase, levels are labeled with the subscript Q. 
Total means the whole history including the two phases or regimes. The present age of the universe 
$10^{61}$, (with $\Lambda = \rho_\Lambda = 10^{-122} = 1/S_\Lambda$) is a {\it lower bound} to the future universe age and similarly 
for the present entropy level $S_\Lambda$. The past  
$10^{-61}$, (with $\Lambda_Q = 10^{122} = \rho_Q = 1/S_Q$ 
is an {\it upper bound} to the extreme past (origin) of the universe and quantum initial entropy, (arrow of time).  
[Similarly, the values given in Fig.1 (in Planck units) for the CMB are the classical CMB age ($3.8 \; 10^5 yr = 10^{57} t_P$) 
and the set of gravitational properties of the universe at this age, and their corresponding precursors in the quantum preceding era 
at $10^{-57} t_P$. $S_\Lambda$ constitute also un upper bound to the entropy of the CMB photon radiation.] 
\begin{figure}
\centering
\centerline{\includegraphics[height=20 cm,width=26cm] {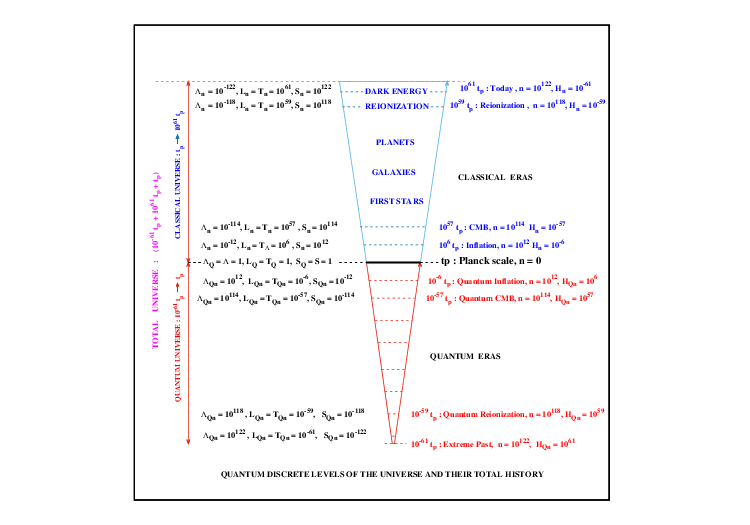}}
\setlength\abovecaptionskip{0cm}\caption{\setlength{
\baselineskip}{0.7\baselineskip}
{\bf The quantum discrete levels of the universe 
from its early trans-Planckian era to today dark 
energy}.  In the pre-Planckian (trans-Planckian) 
phase $10^{-61} t_P \leq t \leq t_P \equiv 
t_{planck}$ the quantum levels are (in 
planck units): $H_{Qn} = \sqrt{(2n + 1)},\; 
\Lambda_{Qn} = (2n + 1),\; S_{Qn} = 1/(2n + 1), n = 
0, 1, 2, ...$, $Q$ denoting quantum. The $n$-levels 
cover {\it all} scales from the past highest excited
trans-Planckian level $n = 10^{122}$, passing the
Planck level $(n = 0)$ and entering the
post-Planckian phase e.g. $n = 1, 
2,...,n_{inflation} = 10^{12},... ,n_{cmb} =
10^{114},...,n_{reoin}=10^{118},...,n_{today} =
10^{122}$. In the post-Planckian universe $t_P \leq
t \leq 10^{61}t_P$ levels are:  $H_{n} = {1}/\sqrt{(2n + 1)}, \Lambda_{n} = 
1/(2n + 1), S_n = (2n + 1)$: as $n$ 
increases,radius, mass and $S_n$ 
increase and {\it consistently} the 
universe {\it classicalizes}. See text 
at the end of Section X.}\end{figure}\\
{\bf Figure (1) caption:} The quantum 
discrete levels of
the universe from its early trans-Planckian era to
today classical vacuum energy (dark energy), namely,
the standard model of the universe completed back in
time with quantum physics in terms of its vacuum
history. The universe is composed of two main
phases: after and before the Planck scale (Planck 
time $t_P$ and Planck units). The complete history 
goes from $10^{-61} t_P$ to $10^{61} t_P$:
 In the pre-Planckian (trans-Planckian) phase
 $10^{-61} t_P \leq t \leq t_P \equiv  t_{planck}$
 the quantum levels are: $H_{Qn} = \sqrt{(2n +
 1)},\; \Lambda_{Qn} = (2n + 1),\; S_{Qn} = 1/(2n +
 1), n = 0, 1, 2, ...$, $Q$ denoting quantum. The
 $n$-levels cover {\it all} scales starting from 
 the past highest excited trans-Planckian level $n 
 = 10^{122}$ with {\it finite curvature} $R_Q =
 10^{122}$, $\Lambda_Q = 10^{122}$ and minimum
 entropy $S_Q = 10^{-122}$, as $n$ decreases:
 $S_{Qn}$ increases, $(H_{Qn}, \Lambda_{Qn})$
 decrease passing the Planck level $(n = 0)$:
 $H_{planck}= 1 =\Lambda_{planck}= S_{planck}$ and 
 entering the post-Planckian phase e.g. $n = 1, 
 2,...,n_{inflation} = 10^{12},... ,n_{cmb} = 
 10^{114},...,n_{reoin}=10^{118},...,n_{today} = 
 10^{122}$. In the post-Planckian universe $t_P 
 \leq t \leq 10^{61}t_P$ the levels are:  $H_{n} = 
 {1}/\sqrt{(2n + 1)},  \Lambda_{n} = 1/(2n + 1), S_n = (2n + 1)$:  As $n$ increases, radius, mass and 
 $S_n$ increase, $(H_n,\Lambda_n)$ decrease and {\it consistently} the universe {\it classicalizes}.  
 The present age of the universe $10^{61} t_P$ with its most classical value $H_{today} = 10^{-61}$, 
 $\Lambda_{today} = 10^{-122} = 1/S_{today}$ is a 
 {\it lower bound} to the future universe age and 
 similarly for the present entropy level $S_n$. The far past $10^{-61} t_P$, (with $\Lambda_Q = 
 10^{122}  = 1/S_Q$) is an {\it upper bound} to the extreme known past ("origin") of the universe and 
 quantum initial entropy, (arrow of time).

\section {Discussion and Clarifications}

A full quantum cosmology dynamics including the full the trans-planckian quantum phase and the discrete time levels is not yet fully accomplished, but that does not mean that this approach does not allow time 
dependence dynamics, on the contrary.
Classical-Quantum Gravity Duality here (CQGD), quantum 
discrete levels in cosmology and its connection to the observational values are a first (non-perturbative)
step towards the completion of a quantum cosmology dynamics including the 
trans-planckian domaine. 

Recall that in the foundation of quantum theory, the quantum 
dynamical equations were written well after that 
classical-quantum duality, wave-particle duality and its implied quantum uncertainty 
were formulated, and they were also motivated by experimental  results. 

As stated in Section IX, after Eqs (9.18)-(9.20), a  direct consequence of the classical-quantum gravity duality (CQGD) relations eg Eqs (2.2), (2.3), (2.4),   and refs [1], [2], [3], is that in the post-Planckian (semiclassical and classical) eras, no trans-Planckian energies are present: only mathematically or artificially (non physical) trans-Planckian energies could be generated in the present universe. The transplanckian energy domain  necessarily remains in the phase  totally before the Plank time  $t_P$, ie in a totally quantum gravity domain. 

A  recent transplanckian Censorship Conjecture (TCC) in string theory eg refs 
\cite{bedrovafa2020} stays that no (low 
energy) effective field theory emerging from superstring theory could lead to a regime 
where fluctuation modes which were initially trans-Planckian ever exit the Hubble radius.  Or equivalently, no modes with four-momenta higher than the Planck scale can enter in 
the low-energy effective action, (thereby 
promoting the role of the Planck mass scale as an 
ultra-violet (UV) momentum cutoff). 
One could says therefore that a part of such TTC finds support in the CQG Duality, even 
if  we have not formulated CQG Duality for such TTC at all,  (and TTC was formulated later than CQGD). And on the other hand, the CQG Duality refers to a ultimate finite 
quantum theory of gravity. In the CQG 
Duality the Planck scale is not a ultimate 
UV cutoff but it is a transition scale (between the dual quantum gravity (transplanckian) and the classical gravity (non transplanckian). The CQG Duality implies a precursor new phase (a whole domain)  before the Planck time,  smaller than the Planck size  which is full transplanckian and the post-planckian universe is necessarily non transplanckian.

The classical-quantum gravity duality here and its quantum levels imply a varying vacuum 
energy ie a varying $\Lambda_n$ at each discrete level $n$ and yield support to it: $H_n$ varies and $\Lambda_n$ varies 
too, even if of course this variation is mild. After all, inflation needs too such a 
vacuum or $\Lambda$ variation. It is known that what it is called Hubble constant is in 
fact a Hubble rate; it should be admitted too that what is called $\Lambda$ cosmological 
constant would be in fact a $\Lambda$ rate.

Recall that the purpose of this approach is the extension of the semiclassical and 
classical cosmological phases to the quantum Planckian and trans-planckian domain, mainly the de Sitter (and quasi de Sitter) early 
and late phases as these are relevant in 
this problem. By no means this is to 
disregard the other expanding intermediate 
phases, although this paper is not 
particularly devoted to them. 

In the Introduction, Eq. (1.1) refers to the 
magnitudes in the post-planckian (classical 
and semiclassical) universe. Eq. (1.2)  (and 
the subscript $Q$)  refers to the magnitudes in the pre-planckian (quantum transplanckian and Planckian) universe. $n = 0$ yields the Planckian (zero point) level (in Planck units). The magnitudes at 
the Planck scale, (the crossing scale), as 
explained in the paper are just a constant: 
their corresponding expressions  (subscript 
$P$)  are totally in terms of the Planck 
constant, as in Eq. (2.3), and for example,  
in the case of the gravitational entropy Eq. 
(2.7), Eq. (3.8).
When we refer to the total or "complete"
magnitudes (also when we stand them  with 
the subscript total or $QG$), we refer to 
the total ($QG$) magnitudes which  are the 
sum of the magnitudes in the three regimes: 
$Q$, $G$ and $P$, and automatically, the 
additive constant magnitude in terms of the 
Planck constant  does appear  ($m_P$ or 
other Planck constant magnitude according to the physical magnitude considered), as for 
example in  Eqs (2.5), (2.6), (3.5), (3.8), 
eg $\Lambda_P$ ,  $\rho_P$  and similarly in the other magnitudes. This applies to all physical 
magnitudes, as described in Sections II , 
III , IV, V.

\medskip

The discrete levels refer to the  quantum space-time levels and in the post-planckian time  eras become the continuous for the classical and semiclassical gravity , but this by no means at all that the QFT of all propagating fields and their interactions (not considered here) became classical: semiclassicasical gravity is precisely QFT in curved continuos space-time.  In the post-planckian epochs the discrete levels becomes classical  for the space-time and gravity, but the genuine QFT of different matter and spins is perfectly valid which is precisely what semiclassical or semiquantum  (non-planckian) gravity does mean. In semiclassical or semiquantum gravity, (eg without the QG variables, without the transplanckian domaine), quantum fields propagate in the classical or semiclassical continuous curved space-time and background fields. Here we are not considering all the different quantum  interactions and contributions to the vacuum although it is possible to consider and separate them but this is not the purpose of this paper.

This approach provides the  results of QFT 
in curved space-time in its own range of 
validity. Semiclassical gravity, eg QFT in 
curved space-times and an effective theory 
of gravity are valid in the post-planckian 
time non quantum gravity universe. In 
particular, in the semiclassical 
or semiquantum gravity and classical gravity regimes, eg  $ H << h_P$, $L <<  l_P$, $\rho 
<<  \rho_P$,  the  expressions from the $QG$ 
variables provide the series in powers of 
$(H/m_P)^2$ plus the constant Planckian 
terms, $h_P$, or $l_P$ or $\rho_P$ 
respectively according to the physical  
magnitude considered: Eqs  (5.6) for 
$H_{HQ}$, Eqs (5.10) for 
$\Lambda_{\Lambda Q}$, Eq (5.21) for 
$\rho_{HQ}$ of this paper. The additive 
constant Planck terms are always present.

This approach does not replace QFT but it 
brings an extension to it, the $QG$ 
variables (and classical-quantum gravity 
duality) are a first step to such (non- 
perturbative) extension. The 
classical-quantum gravity duality thus 
appears as a guiding  property towards the 
construction of a complete theory beyond 
Planck scale. A full complete quantum gravity  theory would be thus a finite theory 
(which is more than a renormalizable 
theory): In the sense of the physical 
renormalization idea, the renormalization 
procedure applies for the non-complete 
theories, because they are valid in an own 
limitated range of validity, and this is so  because such known  QFTs are 
not complete at the Planck scale and beyond 
it.  The extension of QFT in this case is 
by using the complete QG variables which 
provide a finite theory. This is so because 
of the classical-quantum gravity 
duality through the Planck scale. Or, the  non-zero $[X, T]$ space-time commutators due to the related quantum uncertainty, which generate the quantum light-cone, and the zero distance  or UV singularities are smeared out or eliminated. 
(And this 
is not just extending a cut-off  beyond the Planck constant value). The whole 
quantum dynamics linking the total evolution from the early transplanckian phase to 
the present phase is not yet fully known. 

\medskip

Well after the Planck time, during inflation and after it, and in the late universe, the 
semiclassical gravity description does apply,  eg QFT in curved space-time and its back 
reaction. 
Ref \cite{deVegaSanchez2007} and Refs \cite{Sola1}, \cite{Sola2}, \cite{Sola3}
well account for such description.  But is important to keep in mind that 
such QFT gravity and its renormalization is an effective theory, eg it is not the final 
theory: it does not include the planckian and trans-planckian domain, in this sense 
such QFT even if very useful cannot be considered a full quantum cosmology theory. Behind and beyond such effective QFT theory and its renormalized effects,  
it should be a more complete quantum theory from which such effective theory is a sector or approximation. 

In Ref \cite{deVegaSanchez2007} we computed dark energy as the vacuum energy from QFT in the expanding FLRW universe and find  a dark energy equation 
of state $p = w(z) \rho$ in which $w(z) < -1$  asymptotically reaching  the value $-1$ 
from below. Of course, the time dependence comes from the expanding background, and the 
quantum effects (after re-normalization) depend on time.  Also, in Refs \cite{Boya2005}, \cite{Boya2006}, \cite{BDdVS}
we computed the various quantum type corrections to inflation  and to the local renormalized magnitudes, (eg effective potential, correlation 
functions, energy-momentum tensor) from which we found the quantum corrections to the power inflationary spectra.

The present non perturbative approach with the $QG$ variables extends to the quantum higher 
transplanckian phase and goes beyond the perturbative corrections in the power spectra. 
The $QG$ extension and its $H_{QG}$ variables is a first step to cover non-perturbatively both the non-planckian and the quantum (planckian,trans-plankian) gravity domaines. They extend  
(and in particular are in agreement with) the QFT perturbative results of the 
classical/semiclassical (non-planckian) domaine which express themselves as a series in 
powers of $H^2$. For instance, the quantum $QG$ extension of Inflation we computed in Ref [3] and further discuss here below,  Eqs (11.1) and (11.2), agree in 
particular with the quantum corrections to Inflation computed in the framework of  
perturbative QFT in the classical/semiclassical (eg non-planckian) cosmological domaine, Refs \cite{Boya2005}, \cite{Boya2006}, \cite{BDdVS},
\cite{Burigana2010}. We discuss the effective field theory of Inflation in the Ginsburg-Landau approach which is a powerful theoretical scheme for predictions, confrontation to observations and analysis of real data, eg Refs [17]-[20] and Refs therein. This cover a wide class of inflation models, (small field or symmetry breaking families, as well as large field or non-symmetry breaking inflation), not just one model.  The (QG) extended inflationary power spectra are given by

\begin{equation}
[\; \Delta^S_{k,\; QH}\; ] = \frac{[ \;\Delta^S_{k,\; H} \;]}  { [\;1 + (H/h_P)^2\;] \; (\;1 - \delta \epsilon_{QH}\;)^{1/2} }
\end{equation}
\begin{equation}
[ \;\Delta^T_{k,\;QH}\; ] = \frac{ [\; \Delta^T_{k,\; H} \;] }  {[\;1 + (H/h_P )^2\;]}
\end{equation}           

\medskip

QH stands for the total Inflation phase including the known classical/semiclassical Inflation and its precursor: the quantum Inflation era (in the Planckian and trans-Planckian) phase. $[\; \Delta^S_{k,\; H}\; ]$ and $[\; \Delta^T_{k,\; H}\;]$ are the known standard spectra of scalar curvature and tensor perturbations in classical $H$ Inflation, $\delta \epsilon_{QH}$ is the first order QH slow-roll parameter which contains in particular the classical known slow-roll parameter $\epsilon_H $. The total QH spectra contain both: the standard known spectra of the classical/ semiclassical Inflation including its quantum corrections of order $(H/h_P )^2 = 10^{-12}$ in the classical/semiclassical gravity phase $H = 10^{-6} h_P$, at $t = 10^6 t_P$ , (or $10^{-5}M_P$ for the reduced Planck mass $M_P = m_P/\sqrt{8\pi})$, and their quantum dual spectra in the quantum precursor Inflation era $H_Q = 10^6 h_P$ at $t = 10^{-6} t_P$.

\medskip

This description gives support to dynamical vacuum energy e.g dynamical dark energy as 
computed from QFT in a classical FLRW expanding universe, 
Ref \cite{deVegaSanchez2007} and the running vacuum  
Refs \cite{Sola1}, \cite{Sola2}, \cite{Sola3}. The vacuum expectation value of the 
renormalized energy-momentum tensor of quantum fields in a FLRW space-time  (see Ref \cite{deVegaSanchez2007} 
and Refs \cite{Sola1}, \cite{Sola2}, \cite{Sola3}), yields the vacuum energy density and pressure, eg the
dark energy equation of state $p = w(z) \rho$. The results can be expressed in terms of different parametrizations but the different 
magnitudes for the vacuum density, vacuum pressure, dark energy equation of state, as 
well as the inflation fluctuation spectra, all yield to quantum terms which can be 
recasted as a series in powers of $H^2$.  

\medskip

Vacuum dominance at the trans-Planckian era (eg the universe arises from vacuum) implies that this is  
de Sitter or quasi de Sitter phase in the most earliest stage. In the quantum trans-planckian era, from the most  initial  higher  excited  levels, namely $n_{max}$ with 
the smallest  entropies $S_{Qn} = 1/(2n_{max} + 1)$, 
(Section IX), the decreasing or 
de-excitation of the levels through $n_{max}, n_{max} - 1,..., 1$, till $n = 0$, (Planck 
scale), yields the decreasing of $H_{Qn}$ and $\Lambda_{Qn}$, the 
increasing of the size $L_{Qn}$ 
and the increasing of the entropy $S_{Qn}$ passing by its Planck 
value at $n = 0$ and entering the semiclassical and classical phase $S_n = (2n +1)$, $L_n$, $H_n$, $\Lambda_n$ till its most high values in the late universe. The time levels $t_n$ and 
associated physical 
magnitudes are accounting for evolution. These are not the  full dynamical 
wave function equations, but the discrete space time levels and the associated physical 
magnitude levels $H(t_n)$, $\Lambda (t_n)$, $n = 0, 1,2, ...$
consistently account for evolution. 

\medskip

At each level $n$, the physical magnitudes  eg $(L_n, H_n, \Lambda_n, S_n )$ take the 
corresponding value at that time level $t_n$. In the post-Planckian time era $t > t_P$, eg  $(n=0)$, that is energies smaller than Planck energy: the 
levels $n = 1,2, ... $, 
yield increasing times $t_n$ and increasing sizes $L_n$, 
together with increasing 
entropy $S_n$, and smaller Hubble values $H_n$,  $\Lambda_n$, and 
the system becomes more and more classical. In such 
classical/semiclassical evolution, the known QFT 
semiclassical gravity and classical gravity regimes hold, as 
well as  the known QFT in 
curved space-time dynamics, its back reaction  effects, 
and its quantum corrections.

\medskip

The connection between the cosmological constant and the (inverted) harmonic oscillator is derived from the Einstein Equations in Sections VI and VII.   
The space-time discrete $n$ levels and their expressions, $X_n, T_n$  Eqs (8.1) are not a "hypothesis"  nor  a "conjecture" but  derived  expressions.   They are general,  apply to black holes too,  as derived in Ref [2],  also supported by Refs  [1], [3].
$(X_n, T_n)$ is the notation for the (space, time) coordinate levels.  The expressions for $X_n$ and $T_n$ are given by Eq. (8.1)  and the line below it.    

This paper does not treat the cosmic coincidence problem. As is known, a part of the so called cosmological constant problem is connected with the mismatch between the quantum vacuum particle physics estimated value ($10^{122}$) and the low observed value ($10^{-122}$) in Planck units. That is not the total CC problem, but we are mainly interested here in the problem connected with the planckian and transplanckian (that is to say, fully quantum gravity) domains. The problem is to know too which is (or are) the main particle(s) associated to this vacuum and the detection of such particles, as well as the whole evolution. 
And of course, to test such quantum  evolution with the most 
complete cosmological data set. This is important too in  
clarifying or resolving the present $H_0$ problem or tension, as discussed in Refs \cite{Sola1}, \cite{Sola2}, 
\cite{Sola3} with the running vacuum energy.

The cosmological constant (CC) is not really constant along the cosmic history in this approach. This is a difference with a rigid CC but not with dark energy as a vacuum energy. Here the cosmological term appears to be like a "running" vacuum energy with the value of $n$ as $ \sqrt{2n +1}$ (in Planck units) and ultimately with $t$. We are not using a renormalization group approach, $n$ is a space-time level. $n=0$ corresponds to the Planckian constant (or zero point) level. It could be think by analogy as an effective running, although we have not used nor thought $n$ in this way.  In addition, and independently, in ref [49] we have found varying time vacuum energy from QFT in a curved expanding FRWL universe and we have not used analogy with "running", but such time varying could be compared to, or interpreted as, a running.

\medskip

In the process of the classicalization as $n$ increases and 
$t_n = (2n+1)^{1/2}$ increases,  the huge value of the initial 
$\Lambda_n$ diminishes as $\Lambda_n = 1/(2n+1)$, and when $n$ is huge, say $n = 10^{2x}$  with  large $x >> 1$, the 
$\Lambda_n$ value is $10^{-2x}$ smaller than the highly 
quantum initial one,  and hence in the desired range of the  
classical measurement at time $t_n = (2n+1)^{1/2} = 10^x$. 
This is coherently accompassed by  the  decreasing Hubble constant $H_n = 1/(2n+1)^{1/2}$, the 
increasing size $L_n = (2n+1)^{1/2}$,  and 
increasing gravitational entropy $S_n = (2n+1) = 10^{2x}$ from the early eras to 
the present time.

 As explained in the above points, and  Section VIII, Eq.(8.1) and Section IX, $n = 0, 1, 2,...$ is determined by the time 
 levels $t_n = (2n+1)^{1/2}$, and its dual branch $t_{Qn} = 1/(2n+1)^{1/2}$, and conversely. The time, in particular the present time $10^{61}$ (in Planck units $t_P$), determines $n = 10^{122}$,  and therefore the today values for $\Lambda_n = 1/(2n+1)$,  $S_n = (2n+1)$ corresponding to such $n$. That is to say, $n = 10^{122}$ is not an arbitrary choice. 
Of course, any other similar high $n$   corresponding to a time near such era, $n = 10^{100}$ say,  
explains as well the huge difference  between the very early and late
 $\Lambda_n$ values due to the 
classical-quantum duality relations between the transplanckian and the classical (late) eras. 

\medskip

On the other hand, if we would start from the extreme 
early past universe: 
 Is not known what is the most early past remote time, except that it should be a very small fraction of 
 Planck time: If $10^{-x}$ is such at 
 priori unknown number, $x > 0$ to be determined, 
then, starting from  $t_Q = 10^{-x} t_P$, the results of Sections VIII-IX eg $t_{Qn} = 
1/(2n+1)^{1/2}$, yield the quantum level $n = 10^{2x}$,  the most early 
quantum $\Lambda_{Qn} = 10^{2x}$, and  most early 
quantum entropy, (in Planck units)
$S_{Qn} =  10^{-2x}$ . The classical-quantum 
gravity duality relations  yield then for the most late future 
phase observables:   $t_H =10^x t_P$,  $H = 10^{-x}, L_H = 10^{x}$  and  $S_H = 10^{2x}$ for the (dimensionless) gravity entropy.
In particular, today $t_H = 10^{61} t_P$ yields  $x > 61,  t_Q = 10^{-61} t_P$, and  $S_Q = 10^{-122}$ 
as the upper bounds for the most early remote time 
and quantum gravitational entropy; $n = 10^{122}$  
and $\Lambda_Q = 10^{122}$  (in Planck units) are respectively 
the lower bounds for the corresponding quantum level and  energy.  Such quantum huge energies and sizes $L_Q = 10^{-122} << l_P$ are
truly typical of the trans-Planckian 
domain, and  appear in other quantum gravity problems too (as black holes for instance). 
Thus, the early quantum trans-Planckian magnitudes can be 
connected to the late time measurements through the 
classical-quantum duality relations. They provide the most 
stringent upper bounds to the most early remote time and  
early quantum entropy, and most stringent lower bounds for the 
most early quantum level and energy.

\medskip

Let us comment now about the value of $n$ at a given 
time of the cosmological expansion such that the value of 
$\Lambda_n$ does not perturb any segment of the thermal history of 
the universe, e.g. BBN or other known period: Recall that 
$t_n = (2n+1)^{1/2}$, $n = 0, 1, 2,..$ and $\Lambda_n = 1/(2n+1)$ hold for all the post-Planckian cosmic history: $t > 
t_P$, that is in all classical and semiclassical eras, as well 
as all the other gravitational levels $L_n = (2n+1)^{1/2}$, $H_n = 1/(2n+1)^{1/2}$ and $S_n = (2n+1)$ in such-eras.  In the radiation and matter eras, that is in the classical and/or semiclassical gravity regimes, the vacuum dominated 
expressions  for  $H_n$, $\Lambda_n$ and $S_n$ represent  {\it the upper bounds}  (maximum values) to the values of these 
magnitudes in such radiation and matter eras when the vacuum 
energy is not the dominant one, that are computed from the 
semiclassical and classical dynamics (QFT and Einstein 
equations). Recall that $S_n$ is the gravitational 
cosmological entropy, or Bekenstein-Gibbons-Hawking entropy 
and thus this is always un {\it upper bound} to the entropies of the 
different content parts:  radiation, or matter or other partial 
entropies. Therefore, these values will not alter the cosmic 
history, BBN or other part. This does not excludes the 
existence of an early vacuum energy which could explain the $H_0$ 
tensions or even a recently discussed BBN tension Ref \cite{BBN-Pitrou}, 
but we do not discuss such tensions here.

\medskip

This is not an alternative description to 
standard cosmology but a quantum  extension of it, and doing that the vacuum energy appears as 
time evolving. In the QFT description in curved FLRW 
space-time, vacuum energy turns out time dependent, even if such time variation is mild in the late universe. Quantum discrete levels 
of the space-time are neither an alternative description to 
space-time but a more complete description of it.
A time evolving vacuum energy density is not only  a prediction of QFT in curved spacetime but it may provide a 
better description to the cosmological data  than  merely the
constant $\Lambda$ vacuum in the so called
$\Lambda$CDM model see eg, Refs \cite{Sola1}, \cite{Sola2}, \cite {Sola3}. And as it has been
discussed in the literature, phenomenological
models proposing a time-evolving $\Lambda$ (and
hence a dynamical $\rho_{\Lambda}$)  help in
alleviating the several  cosmological parameter
problems or tensions see e.g. Refs 
\cite{DiValentino1}, \cite {DiValentino2} and 
particularly for  $H_0$  \cite{Riess2020}, \cite 
{Riess2021}.

The Standard Model of the Universe in fundamental
grounds is based on General Relativity and 
classical fields and Quantum Field Theory for the
description of matter, and having dark matter and
dark energy in its majoritary components, this
last described by a vacuum energy.  The fact that
the vacuum energy is time varying or time running is 
compatible with semiclassical gravity: General
Relativity and QFT, and is not an alternative to
it, it is neither a modified gravity theory in
the sense of "modified classical Newton gravity"
nor modified classical Einstein  nor other
proposed alternative gravities.

We treat the  area gravitational (Gibbons-Hawking  
\cite{GibbHawkEntropy} and Bekenstein
\cite{bekenstein1981})  entropy in 
terms of the relevant size of the
system  (or object) considered, and its extension
to the quantum gravity transplanckian domaine. 
In the pre-planckian time phase (fully quantum
gravity transplanckian phase), the relevant
appropriate size of the 
quantum system is the Compton length. In the 
post-planckian Universe, the relevant size is the
gravitational size,  the Hubble horizon, the
apparent horizon, as de Sitter gravitational
entropy or 
Gibbons-Hawking entropy. The entropies in the two
different phases are classical-quantum gravity
duals of each other. The total gravitational
entropy is the sum of the 
entropies in the three main gravity regimes:
quantum 
gravity, Planckian and classical/semiclassical 
gravity regimes:   Eq (2.7) of this paper. 
The complete (or QG) variables entail  precisely 
those three regimes, and provide the additive 
constant too, that is  the pure Planckian  scale
term (a constant). This  is discussed in Section
II, Eq (2.7). In  Section II,  the general
physical magnitudes in 
each of the three gravity regimes are explained,
eg Eqs (2.1)-(2.3). Eq (2.4) gives the total QG 
gravitational magnitudes, sum of the quantum
transplanckian (Q), 
classical and planckian ones.
This paper is not discussing already  known
entropy aspects nor the entropy generation
mechanisms or their thermodynamics, eg ref
\cite{solayu2020} and refs therein. The total
entropy in\cite{solayu2020}  refers to the sum of
the area
plus the volume entropy in the post-planckian
time universe (eg  in the 
semiclassical and classical gravity phases), but
the quantum gravity trans-planckian 
dual entropy in the pre-planckian time phase is
not included.

\section{Conclusions}

We have accounted in the introduction and along the paper the main results  and will not include all of them here. We synthetize 
below some conclusions and remarks.

\begin{itemize}

\item{The standard model of the universe is extended back in time with Planckian and trans-Planckian
physics before inflation in agreement with
observations, classical-quantum gravity duality and
quantum space-time. The quantum vacuum energy bends
the space-time and produces a constant curvature de
Sitter background. We find the quantum discrete
cosmological levels: size, time, vacuum energy,
Hubble constant and gravitational (Gibbons-Hawking)
entropy and temperature from the very early
trans-Planckian vacuum to the classical today vacuum
energy. The $n$-levels cover {\it all} scales from
the far past highest excited trans-Planckian level
$n = 10^{122}$ with finite curvature, $\Lambda_Q =
10^{122}$ and minimum entropy $S_Q =  10^{-122}$,
$n$ decreases till the Planck level $(n = 0)$ with
$H_{planck}=1=\Lambda_{planck}= S_{planck}$ and
enters the post-Planckian phase e.g. $n = 1,
2,...,n_{inflation}=10^{12},... ,n_{cmb} =
10^{114},...,n_{reoin}=10^{118},...,n_{today} =
10^{122}$ with the most classical value $H_{today} =
10^{-61}$, $\Lambda_{today} = 10^{-122}$,
$S_{today}=10^{122}$. We implement the Snyder-Yang
algebra in this context yielding a consistent
group-theory realization of quantum discrete de
Sitter space-time, classical-quantum gravity duality
symmetry and a clarifying unifying picture.}

\item{ A picture for the de Sitter background and the  universe epochs emerges,
both for its classical (post-Planckian) and quantum (pre-Planckian) regimes, depicted in Fig (1).
This is achieved by considering classical-quantum gravity duality, 
trans-Planckian physics, quantum space-time and quantum algebra to describe it.
Concepts as the Hawking temperature and the usual (mass) temperature 
are precisely the same concept in the different: classical gravity (post-Planckian) and quantum gravity regimes respectively.  
Similarly, it holds for the Bekenstein-Gibbons and Hawking entropy.}
An unifying clarifying picture emerges 
in terms of the main physical gravitational intrinsic magnitudes of the universe: 
age, size, mass, vacuum energy, temperature, entropy, 
 covering the relevant gravity regimes and cosmological stages: classical, semiclassical, 
quantum Planckian and trans-Planckian eras.
The total or global mass levels are $M_n = m_P \sqrt {2n+ 1}$ \; for all $n = 0, 1, 2, ...$. 
{\it Two}  dual branches $m_{n\pm} = m_P \; [\sqrt{2n +1 } \pm \sqrt{2n}\;]$ do appear for the usual mass variables, covering the
{\it whole mass range}: from the Planck mass ($n = 0$) until the largest cosmological ones in the post-Planckian branch $(+)$, 
and from the smallest masses till near the Planck mass in the pre-Planckian branch $(-)$. 

\item{The quantum space-time structure 
arises from the relevant non-zero space-time commutator $[X, T]$, 
or non-zero quantum uncertainty $ \Delta X \Delta T$.
The {\it quantum light cone} due to the
quantum non-zero uncertainty $[X,T]$ allows a
{\it new quantum region} which is purely quantum
vacuum or zero-point Planckian and trans-Planckian 
energy and constant curvature. 
The quantum de Sitter space-time is described through the relevant quantum non-commutative coordinates and the quantum hyperbolic 
structure. They generalize the classical de Sitter  
 space-time and reduce to it in the classical zero quantum commutator coordinates. Interestingly enough, de Sitter space-time turns out to be {\it discretized} in quantum levels, e.g. $(X_n, T_n) = \sqrt{2n + 1}, n = 0, 1, 2,....$ }

\item {In the post-Planckian  domain, the quantum de Sitter space-time extends in discrete levels
 from the Planck scale level $(n = 0)$ and the quantum (low $n$) levels to the quasi-classical and classical levels (intermediate and large $n$), 
 tending asymptotically for the very large $n$ to a classical continuum space-time. Consistently, these levels have larger gravitational 
 (Gibbons-Hawking) entropy $S_n$, lower vacuum density $\Lambda_n$ and lower Hubble rate $ H_n$. In the pre-Planckian trans-Planckian domain, 
 quantum de Sitter extends 
 from the Planck scale level $(n = 0)$ to the lengths and time smaller than the Planck scale, the quasi-quantum trans-Planckian levels (small and medium $n$), 
 until the deep extreme highly excited trans-Planckian levels (very large $n$) which are those of smaller entropy $S_{Qn}$, higher vacuum density $\Lambda_{Qn}$ and higher $H_{Qn}$.}

\item{Cosmological evolution goes from the pre-Planckian or trans-Planckian quantum phase to the Planck scale and then to the post-Planckian universe: semiclassical accelerated de Sitter era (field theory inflation), then to the classical phase until the present  {\it diluted} de Sitter era. This evolution
between the different gravity regimes could be view as a mapping between asymptotic (in and out) states characterized by the sets
$U_\Lambda$ and $U_Q$, and thus as a Scattering-matrix description: The most early quantum trans-Planckian state 
in the remote past being the "`in-state", and the very late classical dilute state
being the far future or today "`out-state"'.} 

\item{The classical-quantum gravity duality relations are not "abstract" relations: observational values allow to verify them.  In the item here below we include the implications of the trans-Planckian phase for inflation and its effects, which are testable by the CMB and gravitational-wave observations.  Inflation and the fluctuations are derived with this model: It yields (i)  the known (classical/semiclassical)  inflation and its primordial scalar and tensor fluctuation spectra and their predicted numbers tested by the CMB and other cosmological observations, (ii) the quantum corrections and the numbers for these corrections which are in agreement with other independent computations of quantum corrections to inflation (as quantum inflaton decay and inflaton-fermion interactions for instance), (iii) the resummation for them.  All them can be confronted or constrained by the CMB and large scale structure data, as well as in the future by gravitational-wave observations in the primordial gravitational wave frequency range.

The classical-quantum gravity duality relations through 
the Planck scale, are well motivated ones, e.g as shown in \cite{Sanchez2019}, they reduce in particular to the well 
known classical-quantum (de Broglie, Compton) duality 
without gravity.  They are supported by the dynamics of 
quantum fields and strings in curved space-times 
\cite{deVega1994},\cite{deVega1993},\cite{Sanchez2003-1}. 
They are not purely conjectured relations or hypothesis.}

\item{QFT in an expanding universe allows to consider 
particle creation, cosmological perturbations, inflation  
and its nearly scale invariant spectra. In this context, 
QFT in the complete  (with $H_{QH}$) de Sitter space-time 
yields that the total or complete QH inflationary spectra 
turn out expressed by Eq (11.1) and Eq.(11.2) as shown in 
Ref [3] and discussed in Section XI.  
The features the pre-Planckian phase and those the quantum discrete levels could imprint in the inflationary spectra 
deserve to be explored and are beyond the scope of the 
present paper. 
A full quantum description including the quantum 
space-time algebra, its discrete levels and coherent 
states in the complete QH de Sitter group within a group 
theory quantization approach deserve more investigation.}

\item{Inflation is part of the standard cosmological model and is supported by the CMB data of temperature and 
temperature-E polarisation 
anisotropies. This points to $10^{-6} m_P
$, (or $10^{-5} M_P$ for the reduced mass $M_P = 
m_P/\sqrt{8 \pi}$) as the energy scale of Inflation 
\cite{CiridVS},\cite{BDdVS},
\cite{Burigana2010}  
safely below the Planck energy scale $m_P$ of the onset of quantum gravity. This implies that known post-Planckian Inflation is consistently in the 
{\it semiclassical gravity regime}. This 
in turn implies that the preceding phase of Inflation corresponds to a Planckian and pre-Planckian quantum phase. Inflation being a de Sitter, 
(or quasi de Sitter) stage, it has a smooth space-time curvature {\it without any physical space-time singularity}.}

\item{Integrating the above results, and because the earliest stages of the universe are de Sitter (or quasi de Sitter) eras, it appears that there is 
{\bf no singularity} at the universe's origin. First: the so called $t = 0$ Friedman-Robertson Walker  mathematical singularity is 
{\bf not} physical: it is the result of extrapolation of the purely
classical (non quantum) General Relativity theory, {\it out of its domain of physical validity}. The Planck scale is not merely a 
useful system of units but a physically meaningful scale: the onset of quantum gravity, this scale precludes the extrapolation until 
zero time or length. This is precisely what is expected from quantum trans-Planckian physics in gravity: the smoothness of the classical gravitational 
singularities. Second: Inflation (classical or quantum) in the very past 
($10^{6} t_P$ or $10^{-6} t_P$) is mainly a de Sitter or quasi de Sitter smooth constant curvature era {\bf without any curvature singularity}. Third: the extreme past 
 (at $10^{-61} t_P$) is a trans-Planckian de Sitter state of high {\it bounded} trans-Planckian constant curvature and therefore
 {\it without singularity}. This paper is not devoted to the singularity issue but our results here and the whole picture emerging 
 from this paper and \cite{Sanchez3} indicate the trend and insight into the problem.}

\item{Further couplings, interactions and background fields can be added. The conceptual
results here will not change by adding further couplings or interactions,
or further background fields to the background here. Of course, this is just a first input in the 
construction of a complete physical theory and 
understanding {\it in agreement with observations}. 
Besides its conceptual and fundamental physics 
interest, this framework reveals  deep and useful 
clarification for 
relevant cosmological eras and its 
quantum precusors and for the cosmological vacuum. This could provide realist insights 
and science directions where to place the theoretical effort for cosmological missions 
and future surveys such as Euclid, DESI,  WFIRST, LSST-Vera C. Rubin Observatory and Simons Observatory for instance,\cite{Euclid}, \cite{DESI}, \cite{WFIRST},  \cite{LSST}, \cite{Simons}
 and for the searching of cosmological quantum gravitational signals for e-LISA \cite{LISA} for instance, after the 
success of LIGO \cite{LIGO},\cite{DESLIGO} }

\item{The exhibit of $(c, G, h)$ helps in recognizing the different relevant scales and physical regimes.  
Even if a hypothetical underlying "theory of everything" could only require
pure numbers (option three in \cite{Duff}), physical touch
at some level asks for the use of fundamental constants 
\cite{Okun},\cite{Sanchez2003}. Here we used
three fundamental constants, (tension being $c^2/G$). It appears from our study   
here and in ref \cite{Sanchez2019}, that a complete quantum theory 
of gravity would be a theory of pure numbers.}

\item{We can similarly think in quantum string coordinates (collection of point oscillators) 
to describe the quantum space-time structure,
(which is different from strings propagating on a fixed space-time background). 
This yields similar results for the string expectation values 
$ X^2$ and $T^2$ and other related operators  and yields too a quantum  {\it 
hyperbolic space-time width} {\it bending} for the characteristic lines and light cone generators, or for the space-time horizons \cite{Sanchez2019}, \cite{Sanchez2} \cite{Sanchezinprep}.  In string theory the width appears as due to the nonzero size (of the order of the Planck scale) of the quantum string. 
Moreover, the $\sqrt {n}$ quantization we found in this paper is 
like the string mass quantization $M_n = m_s  \sqrt{n}$, $n = 0, 1, ...$
with the Planck mass $m_P$ instead of the string mass $m_s$, that is to say, with the gravitational constant $G/c^2$ instead of the string constant $\alpha'$.} 

\item{Our results on conceptual unification 
ref [15] and QFT and string quantization in 
a wide class of curved space-times e.g refs 
[15], [16], support the classical-quantum 
gravity relations, irrespective of the 
number and nature of the space-time 
dimensions and of whether dimensions be or 
not compactified. The classical-quantum 
gravity duality here does not require the 
existence of any isometry or symmetry in the curved background, neither any other at 
priori condition. Several types of 
relativistic operations e.g. $L \rightarrow 
\alpha' / L $ appear in string theory due to the existence of the dimensional constant 
$\alpha'$ (for example, $T$ - duality, the 
duality symmetry between the winding and 
propagating modes in orbifold 
compactifications). However, the duality we 
are considering, is the classical-quantum 
(or wave particle) duality  (de Broglie or 
Compton type)  relating 
classical/semiclassical and quantum 
behaviors extended to include the quantum 
Planckian and trans-Planckian regime. The de Broglie $L_q = h/p$ or Compton $L_q = h/mc$ 
relation is not the expression of a symmetry transform between physically equivalent 
theories, but a link, through $h$, of two 
different behaviors and regimes. This 
duality and our results on QFT and Quantum 
strings in curved backgrounds inspirated our classical-quantum gravity relations. In a 
similar spirit, $L_Q = l_{pl}^2 /L_G$ and 
more generically, for a general observable 
$O$: $O_Q = o_{Pl}^2/O_G$ relates two 
different classical and quantum gravity 
regimes of Nature through the Planck scale 
(the crossing scale). The complete 
(classical and quantum) magnitudes $O_{QG} = O_Q + O_G = o_{Pl}\;(\;O_G/o_{Pl} + 
o_{Pl}/O_G\;)$  are invariant under the 
exchange $Q \leftrightarrow G$.}

\item{The quantum uncertainty or 
non-commutativity among the space and time 
coordinates acts as a quantum dressing or 
quantum width for the quantum light cone or 
"dressed light cone". In a complete covering of the space-time causal regions, a whole quantum 
vacuum region emerges. In our comments in the 
item above on string theory,  the width appears as due to the non commuting quantum string 
coordinates and the nonzero size of the string.
In the context of QFT gravity, perturbative 
corrections to the dispersion relation, e.g. 
for a scalar field near the light cone is of 
the form $X^2 - T^2 = G/(30\pi)$ \cite{Prokopec}, again as a shifted or quantum corrected light cone 
relation. Recall that quantum back reaction 
effects, gravitational scattering near a event 
horizon structure produces a quantum shift too 
(the shifted horizon) \cite{tHooft},\cite{dVNS1988},\cite{Sanchez1987}.
QFT in curved space-times and their back 
reaction using the complete QG variables (as QH
and its associated variables) is a step to 
describe QFT effects including the 
trans-Planckian domaine and to go beyond 
literature in the field. A full QFT 
description including quantum space-time, 
its discrete levels and coherent states of 
the QH de Sitter group within a group theory approach quantization deserve investigation 
and are beyond the scope of this paper }
\end{itemize}
 
 \medskip

{\bf ACKNOWLEDGEMENTS}

\medskip

The author acknowledges the French National Center of Scientific Research (CNRS)
for Emeritus Director of Research contract, F. Sevre for help with the figure and the 
anonymous referees for useful comments. 
This work was performed in LERMA-CNRS-PSL -Observatoire de Paris-Sorbonne University and 
the Chalonge-de Vega International School Center. 
Discussions and communications with G. 't Hooft, A. Riess 
and R. Penrose are  gratefully acknowledged.

\end{document}